\documentclass{article}

\usepackage{microtype}
\usepackage{graphicx}
\usepackage{subcaption}
\usepackage{booktabs} \usepackage{multirow} \usepackage{array}
\usepackage{rotating} \usepackage{makecell}
\usepackage{pifont}
\newcommand{\cmark}{\textcolor{green!50!black}{\scalebox{1.3}{\ding{51}}}}
\newcommand{\xmark}{\textcolor{red!70!black}{\scalebox{1.3}{\ding{55}}}}

\usepackage{hyperref}
\usepackage{threeparttable}

\usepackage{xcolor}
\usepackage{listings}

\lstdefinestyle{promptstyle}{
  basicstyle=\ttfamily\small,
  breaklines=true,
  breakatwhitespace=true,
  columns=fullflexible,
  keepspaces=true,
  showstringspaces=false,
  tabsize=2,
  backgroundcolor=\color{black!5},
  frame=single,
  rulecolor=\color{black!20},
  framesep=6pt,
  xleftmargin=0pt,
  xrightmargin=0pt
}

\usepackage{xcolor}
\usepackage{pifont}

\usepackage[accepted]{icml2026}

\usepackage{amsmath}
\usepackage{amssymb}
\usepackage{mathtools}
\usepackage{amsthm}

\usepackage[capitalize,noabbrev]{cleveref}

\theoremstyle{plain}

\theoremstyle{definition}

\theoremstyle{remark}

\usepackage[textsize=tiny]{todonotes}

\icmltitlerunning{The Consistency Dilemma in LLMs}

\begin{document}

\twocolumn[
\icmltitle{The Consistency Dilemma in LLMs: \\Generator-Evaluator Agreement and Vulnerability to Mistakes}

\icmlsetsymbol{equal}{*}

  \begin{icmlauthorlist}
  \icmlauthor{Marina Mancoridis}{mit}
  \icmlauthor{Zoë Hitzig}{harvard}   \end{icmlauthorlist}

\icmlaffiliation{mit}{Massachusetts Institute of Technology}
\icmlaffiliation{harvard}{Harvard Society of Fellows}

\icmlcorrespondingauthor{Marina Mancoridis}{marinam@mit.edu}

\icmlkeywords{metamorphic testing, automated evaluations}

  \vskip 0.3in
]

\printAffiliationsAndNotice{}

\begin{abstract}
Large language models are increasingly deployed in agentic pipelines that depend on the model evaluating its own outputs without external verification. The reliability of these pipelines depends on an implicit assumption: that the model applies relevant concepts the same way when it generates an output and later evaluates that output. We propose a new measure, \textit{generator–evaluator self-consistency}, to test this assumption directly and apply it to 10 frontier models across 491 concepts. We find, first, that there is substantial variation in self-consistency. Second, we find that in a clinical setting with physician-validated mistakes \cite{proniakin2025automatic}, across models, those with higher self-consistency are linked to \textit{greater} vulnerability to mistakes. Thus, even when models consistently apply concepts they may not be safe to deploy. This is evidence of a \emph{consistency dilemma} in LLMs: self-consistency is operationally useful, but models that are more consistent are also more prone to mistakes.\footnote{Code and data are available at \url{https://github.com/MarinaMancoridis/ConsistencyDilemma}. The repository includes code to reproduce all main text results, tables and figures, along with a README with replication instructions.}
\end{abstract}

\section{Introduction}

Large language models (LLMs) are increasingly embedded in automated pipelines where they generate outputs and then evaluate, filter, or revise them. Often, the model's second pass is supposed to make the first pass safer or more reliable. A coding agent writes a patch and then reviews its own diff. In synthetic-data generation, a model generates candidate labels and then filters out the ones that appear invalid. A medical assistant explains whether a symptom requires urgent care and then checks whether its advice included the relevant safety warnings. More generally, in agentic pipelines, models play many roles. They produce outputs that rely on some underlying concepts, and then revise, evaluate or otherwise judge the output.

A model's grasp of a concept ought to be consistent across roles. If not, i.e. if the model employs one understanding of a concept while generating an output and another while evaluating it, the outcome of the review step is hard to interpret. The model might flag a sound output against a criterion that played no role in generation, or accept a flawed output because it has stopped enforcing a constraint it applied earlier. This creates a potential source of drift inside self-revising systems---the model may return to the same artifact but no longer treat the underlying problem in the same way when it plays a different role. We therefore need a way to measure whether a model’s use of a concept remains stable when it plays different roles in an automated pipeline. 

We introduce a suite of generator-evaluator tests designed to isolate this form of conceptual self-consistency. Each test is built so that the generator and evaluator outputs should carry a specific relation to each other. To be precise, in all our tests, a model is first asked to produce an output with a specified property, and then the same model is asked to judge whether its output has that property. This lets us measure consistency without needing an external correctness label for the generated artifact. We organize these tests into three families that distinguish qualitatively different ways that concept use can remain consistent or break down: (i) multiple-choice question (MCQ) perturbations, (ii) rationale-based transformations, and (iii) ontological checks. We apply the suite to concepts sourced from saturated general-reasoning benchmarks (MMLU, BBH) and deployment-focused benchmarks (MedicalQA, GDPVal, FinanceQA), and evaluate frontier models across providers. We collect examples of generator-evaluator response pairs, spanning $10$ models, $5$ source benchmarks, and $491$ concepts. We find substantial variation in self-consistency that cannot be explained by performance on the source benchmark questions.

Note that, while standard Q\&A model benchmarks test models' knowledge of concepts, accuracy on a benchmark has little to do with the kind of consistency we aim to measure. A model may answer a benchmark question correctly while failing to reliably answer very closely related questions about the same concept. Take a simple illustration from our results. We ask a model to modify a multiple-choice question from a benchmark so that ``none of the above'' becomes the correct answer, and then ask the same model to solve the modified question. The model may answer the original question correctly, successfully produce the modified question, and then fail to select ``none of the above'' as the answer (see Figure \ref{fig:framework_overview} for an illustration). Nothing about the underlying concept has changed, only the framing of a question about the concept induced by the model itself. In other words, models that are effectively indistinguishable in benchmark accuracy can still diverge in whether they preserve a simple transformation of the task they themselves produced.

So, once we can measure self-consistency, how should we use these measures? One natural idea is to treat this measure as a model-selection criterion, especially in agentic pipelines where drift may be particularly harmful. If two models perform similarly on standard benchmarks, we may prefer to use the one that is more self-consistent. The case for doing so is straightforward. A more consistent model should be easier to use in automated pipelines because its second pass is more likely to be continuous with its first. It should be less likely to drift as it generates, checks, and revises its own outputs. When such a model flags an error on the second pass, the flag reflects a real error rather than an inconsistency in how it applied the concept.

This argument depends on an assumption that consistency does not come with a tradeoff in reliability. But whether consistency actually tracks reliability is an empirical question. So we test, empirically, whether this is the case. Using MedMistakes-Validated, a dataset of clinically realistic vignettes paired with physician-validated model errors, we construct concept-level measures of mistake vulnerability. For each (model, concept) pair, this measure captures how often the model reproduces failure modes that practicing clinicians flagged as genuine errors. The (model, concept) measure of mistake vulnerability is thus an expert-grounded view of reliability that goes beyond benchmark accuracy. And we can relate this measure of mistake vulnerability to self-consistency, also at the (model, concept) level by extracting concepts from the clinical mistakes. 

We find a clear and systematic association between generator–evaluator self-consistency and mistake vulnerability, even after controlling for benchmark performance. Models that are more consistent in how they apply concepts across generation and evaluation are also the ones that reproduce more validated clinical failure modes. The pattern is not driven by a few pathological cases. It reflects a stable ordering across models: those that are high in self-consistency tend to be high in mistake vulnerability, and those that are low in self-consistency tend to be low in mistake vulnerability. Consistency does not translate into reliability in the setting we study. We call this the \textit{consistency dilemma}. 

{The paper proceeds as follows. \Cref{sec:framework} formalizes generator--evaluator self-consistency, and \Cref{sec:taxonomy} introduces a taxonomy of tests. \Cref{sec:pipeline} describes our automated pipeline for instantiating these tests. \Cref{sec:main_suite_results} reports suite-wide results. \Cref{sec:consistency_beyond_accuracy} presents a clinical case study in which the \textit{consistency dilemma} arises. \Cref{sec:related_work} reviews related work, \Cref{sec:limitations} discusses limitations, and \Cref{sec:conclusion} concludes.
}

\section{Framework}
\label{sec:framework}

We propose a measure of self-consistency that captures whether a language model applies the same concept in a stable way across related prompts. To compose this measure, we compare the model’s outputs to each other under concept-linked tests. We define conceptual self-consistency as the rate at which a model's generation and evaluation agree in their invocation of a concept. Specifically, we measure it by asking the model to generate some output that meets a criterion, and then call the model again to evaluate whether its earlier output meets the criterion under which it was generated.  

\textbf{Definition of a `concept'.} In this work, a \emph{concept} is a compact, reusable piece of knowledge that constrains what counts as a valid instance or application across many contexts. A concept can be, for example, a rule, category, or principle. In our setting, each concept is represented by a short label (e.g., \emph{myocardial infarction}, \emph{Bayes' rule}, \emph{drug--drug interaction}). A concept is \textit{not} a one-off fact or a named entity. Nor is it a broad topic.

\textbf{Test structure.} Our measure of conceptual self-consistency is composed through an evaluation suite of paired tests. Each test uses the same model twice: once to \emph{generate} an output that depends on a target concept, and once to \emph{validate} that output under a related prompt that invokes the same concept. The validation step is a consistency check against the model’s own earlier use of the concept.

\begin{figure*}[t]
    \centering
    \includegraphics[width=1.00\textwidth]{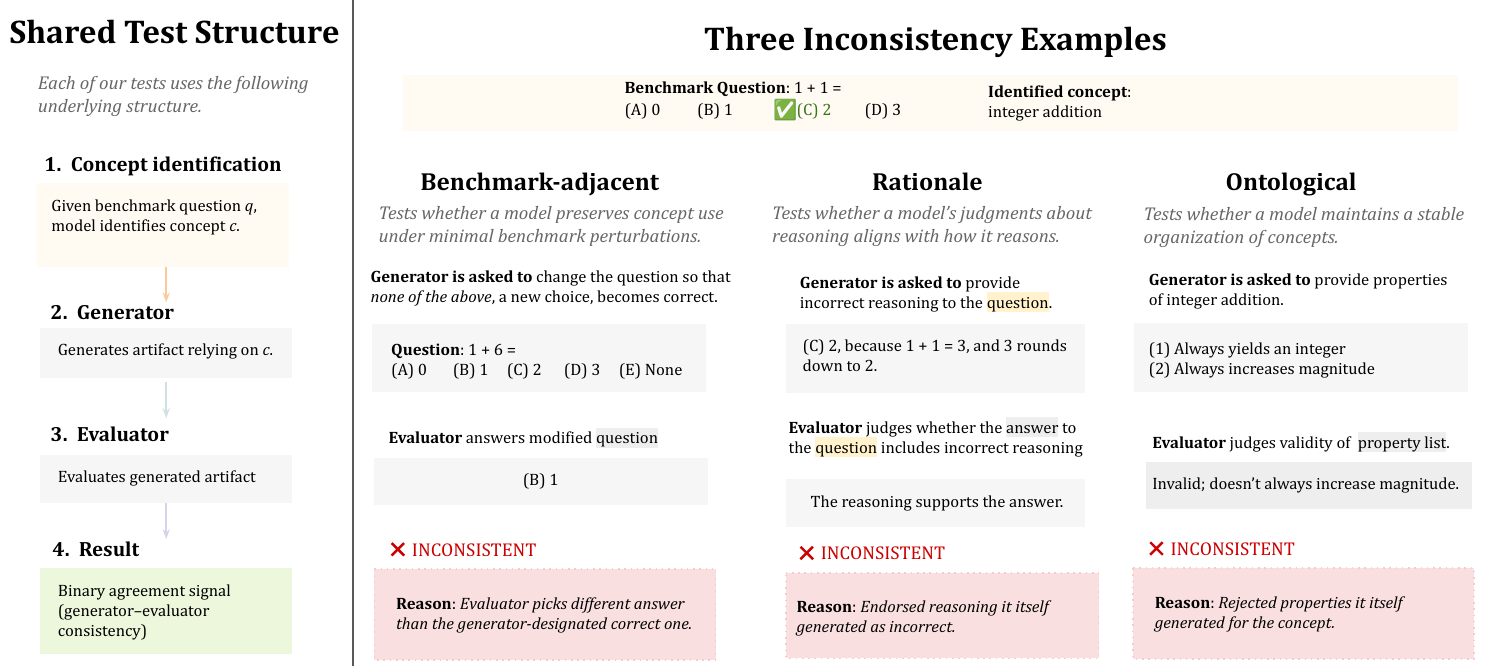}
    \caption{
    \textbf{Generator--evaluator tests for conceptual consistency.}
    \emph{Left:} Shared structure of all tests in our suite: the model identifies a concept in a benchmark question, generates an output relying on that concept, and then evaluates its own output under a related prompt, yielding a binary agreement signal.
    \emph{Right:} One representative test from each of three consistency categories---MCQ perturbation, rationale, and ontological---instantiated using the same benchmark question and identified concept.
    }
    \label{fig:framework_overview}
\end{figure*}

\textbf{Formalization.} Let \( f:\mathcal{X}\to\mathcal{R} \) be a language model that maps a text prompt to a text response. Let \( c \in \mathcal{C} \) be a concept and \( t \in \mathcal{T}(c) \) a test designed to probe use of \( c \). Each test instance \( i \in \mathcal{I}_{c,t} \) consists of paired prompts \((p_g^{(i)}, p_v^{(i)})\). The model generates \(r_g^{(i)} = f(p_g^{(i)})\) from the first prompt and \(r_v^{(i)} = f(p_v^{(i)})\) from the second. A deterministic scoring rule \(s_t:\mathcal{R} \times \mathcal{R} \to \{0,1\}\) assigns a binary agreement label,
\[
Z_{f,c,t,i} = s_t\!\left(r_g^{(i)}, r_v^{(i)}\right),
\]
where \(Z_{f,c,t,i}=1\) if the validation output affirms that the generated output satisfies the criteria specified by test \(t\), and \(0\) otherwise. We interpret \(Z_{f,c,t,i}=1\) as evidence that the model applies the concept consistently within test \(t\) on instance \(i\).

Our contribution is to design a suite of such tests, denoted \( \mathcal{T}(c) = \{t_1, \dots, t_k\} \), each probing a different way in which concept use can remain consistent or break down. Taken together, this suite captures multiple forms of self-consistency, allowing us to analyze patterns across tasks, concepts, and models.

\textbf{Metrics.} Each generator--evaluator test yields a binary outcome. Let
\(Z_{f,c,t,i} \in \{0,1\}\) indicate whether model \(f\) is consistent on the
\(i\)-th instance of test \(t\) for concept \(c\). Let \(n_{c,t}\) denote the
number of instances available for the \((c,t)\) pair. We define the (raw)
consistency rate as
\(\widehat{C}(f,c,t) \;=\; \frac{1}{n_{c,t}} \sum_{i=1}^{n_{c,t}} Z_{f,c,t,i}.\)

Because chance agreement differs across tests (e.g., binary vs.\ multi-choice),
we also report a chance-adjusted score. Let \(p_{\mathrm{ch}}(t)\) denote the
chance-level agreement implied by the structure of test \(t\) (e.g., \(1/2\) for
binary tests; \(1/K\) for selecting a specific option among \(K\) choices). For details on the computation of \(p_{\mathrm{ch}}(t)\) in our suite, see Appendix Section \ref{app:chance_correction}. We define
\(\widehat{C}^{\mathrm{adj}}(f,c,t)
\;=\;
\frac{\widehat{C}(f,c,t)-p_{\mathrm{ch}}(t)}{1-p_{\mathrm{ch}}(t)}.\) Note that \(\widehat{C}^{\mathrm{adj}}\) can be negative, indicating worse-than-chance
agreement.

We additionally aggregate by (i) averaging over concepts to obtain a test-level
score \(\widehat{C}(f,t)\), (ii) averaging over models to obtain a concept-level
score \(\widehat{C}(c)\), and (iii) averaging over test groups in our taxonomy
to obtain group-level scores \(\widehat{C}(f,\mathcal{G})\) (and analogously for
\(\widehat{C}^{\mathrm{adj}}\)).

Figure~\ref{fig:framework_overview} illustrates the generator--evaluator structure shared by all tests in our suite, along with three representative consistency types. In each test, the same model first generates an output that relies on a target concept and then evaluates that output under a related prompt invoking the same concept. The right panel shows how inconsistency can arise in qualitatively different ways, even when the underlying concept and benchmark question are held fixed. These examples motivate our taxonomy of MCQ perturbation, rationale, and ontological consistency, which we formalize in the following section.

\section{A Taxonomy of Conceptual Consistency}
\label{sec:taxonomy}

Our evaluation suite consists of automated tests designed to measure \emph{self-consistency} in concept use. All tests share the same generator--evaluator structure (as defined in Section~\ref{sec:framework}), but they probe different types of consistency. We group the tests into three categories: \emph{MCQ perturbation}, \emph{rationale-based}, and \emph{ontological}.

\textbf{MCQ-perturbation self-consistency.} These tests probe behavior under small, controlled edits to a multiple-choice question (MCQ). The model first produces a perturbed MCQ and is then asked to solve it. For example, the model can first be asked to modify a multiple-choice question so that none of the original answer choices are correct and a new ``None of the above'' option becomes correct, and can then be asked to solve the modified question to assess whether it correctly selects the added option. Other work has shown models to be accurate on the original question yet behave inconsistently under controlled changes \cite{hendrycks2021measuring, srivastava2023beyond, mancoridis2025potemkin}.

\textbf{Rationale self-consistency.} These tests measure whether a model can audit its own reasoning traces consistently across linked prompts. In the generator step, the model produces an answer together with a rationale under a specified condition (e.g., a fully correct explanation, an incomplete one, or an intentionally flawed one). In the evaluator step, the \emph{same model} is asked to judge whether the answer--rationale pair satisfies the relevant criterion (e.g., whether the rationale is valid, complete, or actually supports the answer). This is operationally important in agentic revision and LLM-as-a-judge pipelines, where models are increasingly used as automated evaluators \cite{zheng2023judging}: if a model produces reasoning that does not support its own conclusion (self-contradictory reasoning) and fails to flag it, the pipeline can systematically rubber-stamp flawed explanations \cite{liu2024self}.

\textbf{Ontological self-consistency.} Some tasks require models to maintain a coherent organization of concepts and their relationships. Concepts are defined in part by how they relate to categories, instances, properties, and constraints, and stable use depends on applying these relationships consistently across contexts. This mirrors findings that models can perform well on world-model diagnostics while producing failures on related but subtly different tasks \cite{vafa2024evaluating}. Concretely, the ontological tests ask the model to generate examples, non-examples, attributes, or properties and then validate them under a related prompt.

\textbf{Scope of the taxonomy.}
Our goal is not to enumerate every possible self-consistency check, but to capture a set of qualitatively distinct failure modes. Models can be highly self-consistent in one category and unstable in another. A complete table of tests and their descriptions appears in Appendix~\Cref{sec:app_test_schematics}, and the full generator and evaluator prompts are provided in Appendix~\Cref{sec:app_prompts}.

We now describe our automated evaluation suite design. 

\section{Evaluation Pipeline}
\label{sec:pipeline}
\textbf{Overview.}
We operationalize self-consistency using a fully automated, black-box pipeline. The basic unit of analysis is an \emph{evaluation instance}: one model, one benchmark question, one extracted concept label, and one self-consistency test. In each instance, the same model plays both roles (generator and evaluator). Each instance yields a binary agreement label \(Z\) (Section~\ref{sec:framework}), and we aggregate these labels across instances to compute self-consistency scores.

\textbf{Benchmarks as concept sources.} MCQ-perturbation and rationale tests require a benchmark question as context. We draw questions from two classes of benchmarks: saturated general-reasoning benchmarks (\textbf{MMLU} \cite{hendrycks2021measuring}, \textbf{BBH} \cite{srivastava2023beyond}) and mission-critical benchmarks (\textbf{MedicalQA} \cite{hosseini2024a}, \textbf{GDPVal} \cite{patwardhan2026gdpval}, \textbf{FinanceQA} \cite{mateega2025financeqa}).

Within each benchmark, we first randomly sample a fixed set of question slots before any model-specific steps. These slots are shared across models, so each model is evaluated on the same underlying questions; if a later prerequisite fails for a given model (e.g., no usable concept label or MCQ format), we \emph{gate out} that model--question pair and record it as non-scoring rather than replacing it with a new question.

\begin{table*}[t]
\centering
\begin{threeparttable}
\scriptsize\fontsize{6.5}{9}\selectfont
\setlength{\tabcolsep}{2.4pt}
\renewcommand{\arraystretch}{1.05}
\begin{tabular}{llcccccccccc|c}
\toprule
\textbf{Type} & \textbf{Test} & \textbf{Opus 4.5} & \textbf{Sonnet 4.5} & \textbf{DeepSeek} & \textbf{Gemini 2.5} & \textbf{Gemini 3} & \textbf{GPT-4o} & \textbf{GPT-5} & \textbf{GPT-5.1} & \textbf{GPT-5.2} & \textbf{Mistral} & \textbf{All} \\
\midrule
\multirow{2}{*}{MCQ-P.}
& None of the Above & 0.31 (0.21) & 0.50 (0.19) & 0.17 (0.20) & 0.71 (0.15) & 0.79 (0.13) & 0.11 (0.21) & 0.79 (0.13) & 0.20 (0.16) & 0.13 (0.16) & 0.12 (0.18) & \textbf{0.40 (0.06)} \\
& Previous Answer & 0.47 (0.21) & 0.62 (0.16) & 0.43 (0.18) & 0.26 (0.16) & 0.75 (0.13) & 0.38 (0.17) & 0.53 (0.15) & 0.42 (0.17) & 0.47 (0.17) & 0.14 (0.17) & \textbf{0.45 (0.05)} \\
\midrule
\multirow{4}{*}{Rat.}
& Right for Right Reason & 0.68 (0.17) & 0.16 (0.18) & 0.63 (0.13) & 0.81 (0.09) & 0.80 (0.09) & 0.84 (0.09) & 0.82 (0.08) & 0.59 (0.12) & 0.44 (0.14) & 0.72 (0.13) & \textbf{0.66 (0.04)} \\
& Right for Wrong Reason & 0.33 (0.21) & 0.24 (0.18) & 0.65 (0.13) & 0.85 (0.08) & 0.90 (0.07) & 1.00 (0.00) & -0.05 (0.16) & 0.56 (0.12) & 0.65 (0.16) & 0.47 (0.16) & \textbf{0.58 (0.04)} \\
& Wrong for Wrong Reason & 0.47 (0.20) & 0.74 (0.12) & 0.67 (0.12) & 1.00 (0.00) & 0.95 (0.05) & 0.89 (0.07) & 0.59 (0.14) & 0.91 (0.06) & 0.81 (0.13) & 0.57 (0.16) & \textbf{0.79 (0.03)} \\
& Correct for Incomplete Reason & -1.00 (0.00) & -0.55 (0.15) & 0.73 (0.11) & 0.32 (0.15) & 0.23 (0.16) & 0.90 (0.07) & -0.68 (0.11) & -0.67 (0.11) & -0.71 (0.12) & 0.29 (0.18) & \textbf{-0.08 (0.05)} \\
\midrule
\multirow{6}{*}{Ont.}
& Valid Example & 0.91 (0.09) & 0.69 (0.13) & 0.84 (0.09) & 0.96 (0.04) & 1.00 (0.00) & 0.90 (0.07) & 0.96 (0.04) & 0.96 (0.04) & 0.86 (0.08) & 0.64 (0.14) & \textbf{0.89 (0.02)} \\
& Invalid Example & 0.74 (0.14) & 0.69 (0.13) & 0.59 (0.13) & 0.52 (0.13) & 1.00 (0.00) & 0.85 (0.08) & 0.91 (0.06) & 0.70 (0.11) & 0.69 (0.11) & -0.44 (0.18) & \textbf{0.67 (0.04)} \\
& List True Attributes & 0.82 (0.12) & -0.23 (0.17) & 0.95 (0.05) & 0.74 (0.10) & 0.81 (0.09) & 1.00 (0.00) & 0.82 (0.08) & 0.96 (0.04) & 0.91 (0.06) & 0.71 (0.13) & \textbf{0.78 (0.03)} \\
& List Wrong Attributes & 0.91 (0.09) & 1.00 (0.00) & 0.84 (0.09) & 0.95 (0.04) & 1.00 (0.00) & 0.80 (0.09) & 1.00 (0.00) & 0.91 (0.06) & 0.96 (0.04) & 0.63 (0.15) & \textbf{0.91 (0.02)} \\
& List True Examples & 0.42 (0.19) & -0.48 (0.16) & 0.71 (0.12) & 0.77 (0.10) & 0.80 (0.09) & 0.75 (0.10) & 0.70 (0.11) & 0.88 (0.07) & 0.55 (0.13) & 0.52 (0.16) & \textbf{0.60 (0.04)} \\
& Has Property & 0.82 (0.12) & 0.00 (0.18) & 0.94 (0.06) & 0.81 (0.09) & 1.00 (0.00) & 0.91 (0.06) & 0.59 (0.12) & 0.87 (0.07) & 0.68 (0.11) & 1.00 (0.00) & \textbf{0.77 (0.03)} \\
\bottomrule
\end{tabular}
\end{threeparttable}
\caption{\textbf{Generator--evaluator agreement by model and test type}. Reported values are chance-corrected agreement rates, normalized so that 0 corresponds to chance performance, with standard errors in parentheses.}
\label{tab:model_by_test_chance}
\end{table*}

\textbf{Concept identification.}
Because our tests are concept-linked, the pipeline attempts to identify a single concept for each pre-sampled question. For each \((\text{model},\text{question})\) pair, we ask the model whether the question relies on an underlying concept and, if so, to provide a short concept label. If no single usable concept is identified, the corresponding instance is marked as excluded from scoring (a gate) rather than replaced by a new question. The concept label extracted by the model is then used across all self-consistency tests applied to that question. The full concept-extraction prompt is provided in Appendix~\Cref{sec:app_concept_extraction}.

\textbf{Models.}
We evaluate ten frontier large language models:
\textbf{GPT-4o}, \textbf{GPT-5}, \textbf{GPT-5.1}, and \textbf{GPT-5.2} (OpenAI);
\textbf{Claude Opus 4.5} and \textbf{Claude Sonnet 4.5} (Anthropic);
\textbf{Gemini 2.5 Pro} and \textbf{Gemini 3 Pro Preview} (Google);
\textbf{DeepSeek Chat} (DeepSeek); and
\textbf{Mistral Large} (Mistral AI). This range of models allows us to assess how patterns of conceptual consistency vary by model.

\textbf{Applying the consistency tests.} Each evaluation instance applies one self-consistency test from the taxonomy (Section~\ref{sec:taxonomy}) and records a binary agreement label $Z$ obtained by deterministic parsing of the evaluator’s response.

\textbf{Validity conditions.}
Some tests are only meaningful when prerequisites are satisfied (e.g., multiple-choice structure, baseline benchmark correctness, or successful concept identification). We record these prerequisites as explicit scoring gates and exclude failing instances from self-consistency aggregates while retaining them in logs. Gate definitions and activation rates appear in Appendix~\Cref{sec:app_gates}, and a discussion of what these gates imply for our analyses appears in \Cref{sec:limitations}.

To further assess whether instruction-following failures could confound our consistency measurements, we conduct a targeted audit of instruction-following well-formedness using an automated verifier, supplemented by manual review. The results, reported in Appendix~\ref{sec:app_instruction_following}, indicate that malformed outputs are rare and do not drive the consistency patterns observed in the main analysis.

\section{Suite-Level Patterns of Self-Consistency}
\label{sec:main_suite_results}
In this section, we summarize patterns in self-consistency across models, benchmarks, and concepts. The dataset contains $3{,}951$ evaluation instances; each instance applies one test \(t\) to one source benchmark question \(q\) for one model \(f\), yielding a binary generator--evaluator agreement outcome. In total, we evaluate $10$ models on questions drawn from $5$ source benchmarks; concept extraction yields $491$ distinct concepts used to instantiate tests.

\textbf{Generator–evaluator agreement.} Table~\ref{tab:model_by_test_chance} reports generator–evaluator agreement for each model and test after correcting for test-specific chance baselines. Because tests differ in structure (e.g., binary judgments versus multiple-choice decisions), raw agreement values are not directly comparable across tests. We therefore normalize each test so that $0$ corresponds to chance-level behavior and $1$ to perfect agreement. Under this normalization, positive values indicate agreement above chance, values near zero indicate chance-level behavior, and negative values indicate performance worse than chance. Details of the chance baselines and normalization procedure appear in Appendix~\ref{app:chance_correction}, and uncorrected agreement rates are reported in Appendix~\ref{app:raw_agreement_table}.

We see that self-consistency varies systematically across models and test types. For instance, Gemini~3 and GPT-4o achieve near-ceiling agreement on several ontological tests, while Claude Sonnet 4.5 performs substantially worse on closely related tasks.

\begin{figure}[h]
\centering
    \includegraphics[scale = .35]{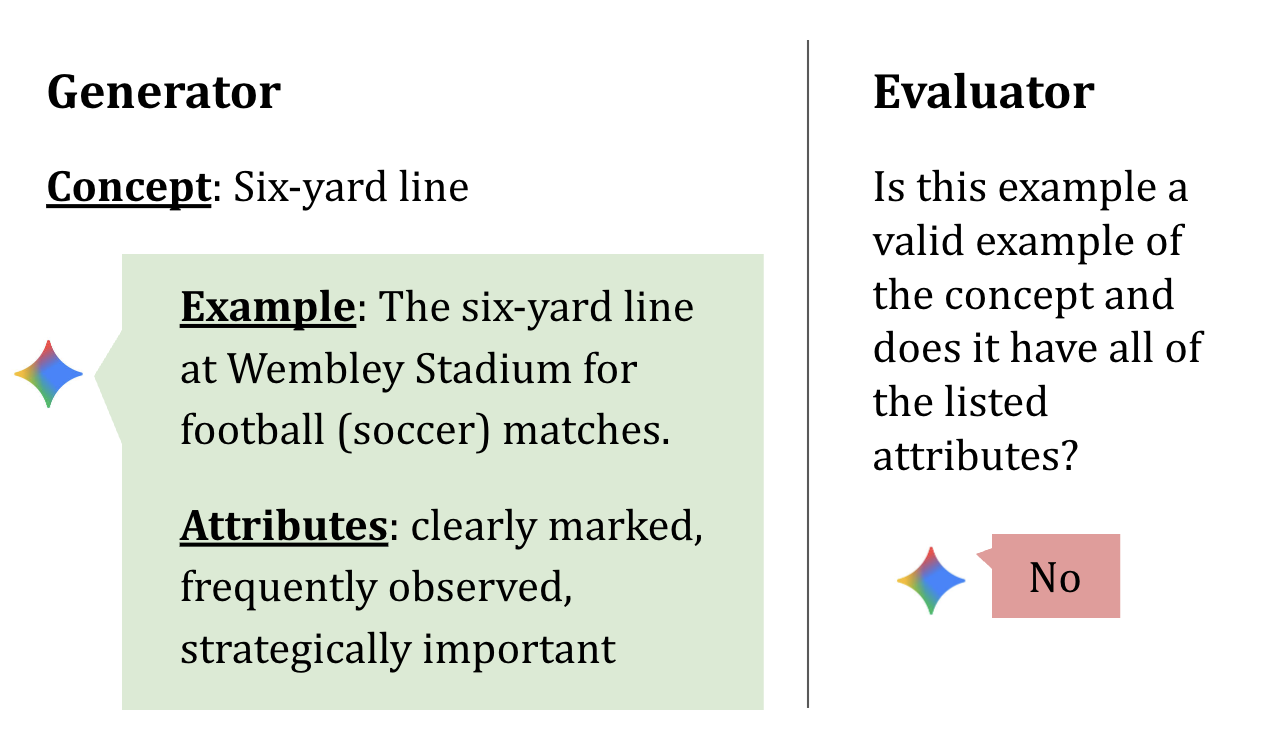}
    \caption{\textbf{An example of a consistency failure on the ``List True Attributes" (ontological) test}, from Google's Gemini-2.5 model.}
    \label{fig:failure_example}
\end{figure}

\textbf{Qualitative failure examples.}
\label{sec:results_examples}
Figure~\ref{fig:failure_example} shows an instance of an ontological consistency failure. The model generates an example–attribute pair describing the six-yard line at a soccer stadium and lists attributes such as being clearly marked, frequently observed, and strategically important. When asked to evaluate this output, the model rejects it, indicating that the generated example does not satisfy all listed attributes.

\textbf{Robustness of generator--evaluator agreement.}
We assess the stability of generator--evaluator agreement under repeated evaluation. We sample $N=200$ instances stratified by test type, and re-run each instance $K=5$ times using identical prompts and decoding settings. For each instance, we compute the fraction of runs that agree with the modal judgment.

Agreement outcomes are highly stable (mean $0.923$). Stability is slightly higher when the modal outcome is agreement than when it is disagreement (difference in means $=0.056$; 95\% bootstrap CI $= [0.003, 0.113]$). Additional robustness analyses are reported in Appendix~\Cref{sec:app_robustness}.

\section{Self-consistency is associated with mistake vulnerability in deployment}
\label{sec:consistency_beyond_accuracy}

Over the preceding sections, we measured self-consistency as a behavioral property of how models apply a concept across related prompts. We now ask what that property implies in a deployment-relevant stress test: given a fixed catalogue of clinician-validated, high-stakes failure modes, how often does a model reproduce the same failure when the clinical situation is presented in superficially different forms? Using MedMistakes-Validated \cite{proniakin2025automatic}, we quantify concept-level vulnerability to these known errors and relate it to self-consistency, conditional on benchmark medical knowledge.

Our outcome is \emph{mistake vulnerability}: for each model--concept pair, we measure how often the model reproduces validated mistakes associated with that concept. 

\begin{figure*}[t!]
    \centering
    \includegraphics[width=0.95\textwidth]{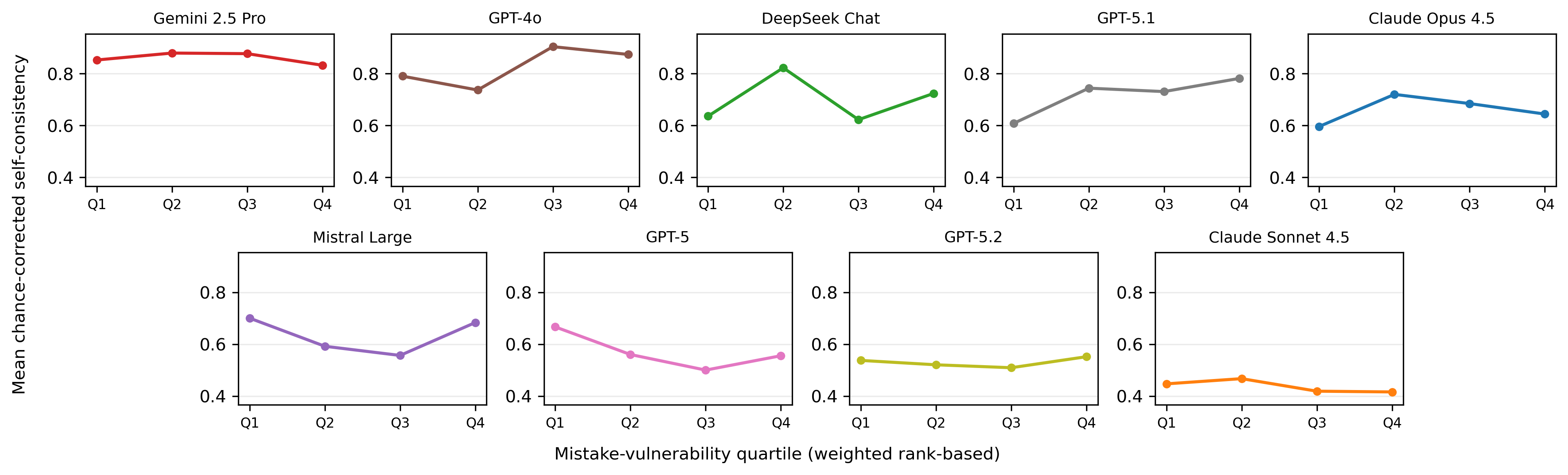}
    \caption{\textbf{Model (chance-corrected) self-consistency profiles across mistake-vulnerability quartiles.} We bucket \((\text{model},\text{concept})\) pairs into quartiles by mistake vulnerability \(Y^{\ast}_{m,c}\) using weighted ranks (so each quartile contains approximately one quarter of the total effective mistake mass). Within each quartile, we plot each model’s weighted mean self-consistency.}
    \label{fig:vulnerability_profiles}
\end{figure*}

\textbf{Quantities used in the medical analysis.} We construct three quantities at the $(m,c)$ level (full details of the data construction are in Appendix~\ref{sec:app_full_process}
):

 \textit{Mistake vulnerability (MedMistakes).}
    For each model and each validated mistake, the dataset provides a set of clinician-provided tags and a binary indicator for whether each model reproduces the mistake. We treat tags as candidate medical concepts and focus on a curated concept set (Appendix~\Cref{sec:app_concept_selection}). To avoid double-counting mistakes with multiple tags, we construct a weighted concept-level \textit{mistake vulnerability} outcome \(Y^{\ast}_{m,c}\in[0,1]\) that splits each mistake evenly across its tags, so each underlying mistake contributes total weight of one across concepts (Appendix~\Cref{sec:app_medmistakes_weighted}). That is, the weighting prevents a multi-concept mistake from being counted multiple times in a concept-level outcome.

  \textit{Benchmark performance (MedQA).} To control for baseline medical knowledge, we measure benchmark performance using MedQA~\cite{jin2021disease}. For each \((m,c)\), \(A_{m,c}\in[0,1]\) is the model’s accuracy on MedQA questions associated with concept \(c\). We find that the benchmark performance is high: $A_{m,c}$ averages 0.90 (SE = 0.005).
    
\textit{Self-consistency.} We compute \(C_{m,c}\) as the average chance-corrected generator--evaluator agreement across the $12$ concept-linked tests, normalizing each test so that 0 corresponds to chance-level agreement and 1 to perfect agreement (Appendix~\Cref{app:chance_correction}). {Appendix~\Cref{app:consistency_distributions} additionally reports the distribution of self-consistency scores across all \((m,c)\) pairs and separately by test}.

\textbf{Statistical specification.} We construct a dataset with one row per \((m,c)\) pair. Let \(Y^{\ast}_{m,c}\) denote mistake vulnerability (MedMistakes), \(A_{m,c}\) denote benchmark performance (MedQA), and \(C_{m,c}\) denote self-consistency. The outcome \(Y^{\ast}_{m,c}\) lies in \([0,1]\), while \(A_{m,c}\in[0,1]\) and \(C_{m,c}\in[-1,1]\).

Because \(Y^{\ast}_{m,c}\) is fractional, we estimate a weighted \emph{fractional logit} model~\cite{papke1996econometric}:
\[
\mathbb{E}\!\left[Y^{\ast}_{m,c}\mid A_{m,c}, C_{m,c}\right]
\;=\;
\mathrm{logit}^{-1}\!\left(\beta_0 + \beta_1 A_{m,c} + \beta_2 C_{m,c}\right),
\label{eq:coh_flogit}
\]
where each \((m,c)\) observation is weighted by the number of mistakes that contribute to $Y^*_{m,c}$, with standard errors clustered at the concept level. We weight observations by the effective number of validated mistakes contributing to each \((m,c)\) outcome (the sum of tag-splitting weights), so concept pairs supported by more mistakes receive more weight.

Our hypothesis of interest is whether self-consistency is associated with mistake vulnerability after controlling for medical benchmark performance i.e. whether $\beta_2 \neq 0.$

\textbf{Results.}  
Across \(N=440\) $(m,c)$ observations, chance-corrected self-consistency is positively associated with mistake vulnerability, conditional on benchmark performance. In our deployment-weighted fractional logit, the estimated
coefficient is $\hat\beta_2 = 1.116$. With standard errors clustered on
concept, this association is highly significant ($\mathrm{SE} = 0.224$,
$p < 10^{-6}$). Because the variation in self-consistency that identifies
$\hat\beta_2$ runs primarily across models, we also report two-way clustered
standard errors on concept and model \citep{cameron2011robust}---the
association remains significant at the 5\% level ($\mathrm{SE} = 0.533$,
$p = 0.036$). Intuitively, benchmark performance is negatively associated with mistake
vulnerability ($\hat\beta_1 = -1.732$, $p = 0.042$ under two-way clustering).

The relationship between consistency and mistake vulnerability is one that holds between models, not within a single model. In other words, we find that higher self-consistency tracks higher vulnerability, yet within a given model it does not predict which concepts are most vulnerable. When we add model fixed effects to the regression specification above,
the coefficient on self-consistency falls to $\hat\beta_2 = 0.147$
($\mathrm{SE} = 0.334$, $p = 0.66$). That is, once we condition on the model, how self-consistent it is on a given concept does not
predict where it is more vulnerable to mistakes. We therefore read
self-consistency as a model-level behavioral signature that is informative for
comparing and selecting models. 
Full implementation details and regression results, including the
fixed-effects specification, are in Appendix Sections
\ref{sec:app_medmistakes_construction} and \ref{sec:app_cc_results_full}.

{\textbf{What kind of consistency matters in diagnostic scenarios?} We also decompose the aggregate consistency score to see which test components account for the consistency–vulnerability association. Recomputing chance-corrected consistency while leaving out each test yields a stable relationship in nearly all cases, with one clear exception: removing \emph{Correct for Incomplete Reason} (CFIR) substantially attenuates the effect. CFIR exhibits relatively large cross-model variation (see \Cref{tab:model_by_test_chance}), so it has more leverage to explain differences. Full results are reported in Appendix~\Cref{sec:app_loo_cc}.}

To further probe the consistency–vulnerability link, we zoom in on a single concept: \textit{triage}. We compare how models behave on our self-consistency suite with the kinds of clinician-validated failures MedMistakes associates with triage. In particular, we focus on Correct for Incomplete Reason (CFIR), which tests whether a model will generate and then endorse an answer supported by under-specified reasoning. This mirrors a common triage failure in MedMistakes: models point in the right general direction but omit critical escalation criteria, safety warnings, or follow-up steps. See Appendix~\ref{sec:app_concept_trace_triage} for the full output of one  \emph{Correct for Incomplete Reason} (CFIR) test and the corresponding MedMistakes failure for the concept \emph{triage}.

\textbf{Model-level consistency profiles.}
At the model level, is consistency stable across mistake-vulnerability levels? To answer this, we stratify $(m,c)$ pairs into quartiles of mistake vulnerability $Y^*_{m,c}$ using weighted ranks (weights given by the effective mistake mass behind each cell). Within each quartile, we compute each model's weighted mean self-consistency. Figure~\ref{fig:vulnerability_profiles} shows that the relative ordering of models is largely preserved across vulnerability quartiles: models that are more self-consistent in low-vulnerability regions remain more self-consistent in high-vulnerability regions. This stability suggests that self-consistency behaves like a model-level behavioral signature.

\begin{figure*}[t]
    \centering
    \includegraphics[width=0.95\textwidth]{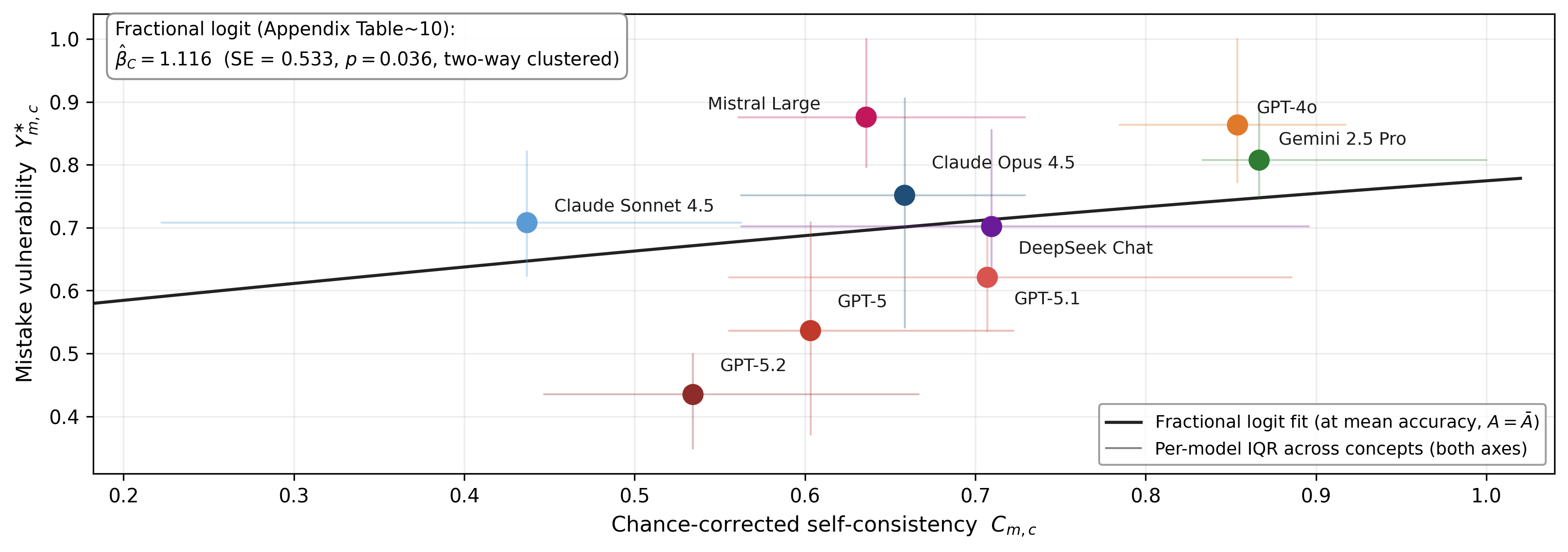}
\caption{\textbf{Association between self-consistency and mistake vulnerability across models.} 
Each point represents a model, positioned by its mean chance-corrected self-consistency \(C_{m,c}\) (x-axis) and mistake vulnerability \(Y^{\ast}_{m,c}\) (y-axis), aggregated across concepts. Horizontal and vertical bars denote the interquartile range across concepts for each model on the respective axis. The solid line shows the fitted fractional logit relationship evaluated at mean accuracy (\(A=\bar{A}\)); the inset reports the estimated coefficient, clustered standard error, and \(p\)-value. Models with higher self-consistency tend to exhibit greater vulnerability to validated mistakes, as indicated by $p < 0.05$ in our regression.}
    \label{fig:consistency_trap_quadrants}
\end{figure*}

\textbf{Consistency--vulnerability association.} {Figure~\ref{fig:consistency_trap_quadrants} provides a high-level visualization of heterogeneity across models in the relationship between self-consistency and mistake vulnerability. Each point corresponds to a model, positioned by its weighted mean chance-corrected self-consistency and weighted mean mistake vulnerability across concepts using mistake-mass weights. The figure shows that models differ substantially along both dimensions: some models exhibit relatively low self-consistency and low vulnerability to validated mistakes, while others are both highly self-consistent and highly vulnerable. The interquartile bars around each point show the spread across concepts within each model, illustrating that within-model variation is large relative to the differences between models.
 We overlay the fractional logit relationship evaluated at mean benchmark accuracy as a solid line on these axes for reference (note that the solid line in Figure \ref{fig:consistency_trap_quadrants} \textit{is not} a regression fit to the model-level points shown).}

\section{Related Work}
\label{sec:related_work}

We study generator--evaluator consistency: whether a model’s outputs remain mutually consistent across related prompts in the absence of a correctness oracle. 

Our work is closely related to recent studies of \emph{LLM-as-a-judge} evaluation, especially work documenting self-preference and correlated reasoning failures across models. Prior studies show that LLM evaluators systematically favor their own generations or stylistic patterns \cite{panickssery2024llm, wataoka2024selfpreference}, while others demonstrate that frontier models often make highly correlated mistakes and converge on similar erroneous reasoning processes \cite{kim2025correlated, goel2025great, jiang2026artificial}. These findings raise concerns that agreement among models or between generation and evaluation stages may reflect shared biases rather than independent verification. Our work differs in a few key ways. First, unlike prior approaches that rely on \emph{cross-model evaluation} or pairwise model comparison, we study consistency entirely \emph{within a single model}’s generator--evaluator loop. Second, rather than measuring agreement at the level of full outputs or preferences, we operate at the level of reusable \emph{concepts} and test whether the same underlying rule or principle is applied consistently across related contexts. Third, we connect these consistency patterns to downstream deployment failures in a clinician-validated medical setting. See Appendix Section \ref{app:comparison_table} for a tabular comparison of these works.

Automated consistency checks like ours are one approach to the broader ``oracle problem'' in software testing and machine learning, where ground-truth labels are unavailable or prohibitively expensive to obtain \cite{barr2014oracle}. One class of such checks is \emph{metamorphic testing}, which replaces correctness labels with expected relationships between outputs under controlled input transformations; it has been adapted to LLMs and NLP pipelines to surface inconsistencies without labeled data \cite{cho2025metamorphic, ribeiro2020beyond}.

Many types of metamorphic relations have been studied in language systems. \citet{li2024empirical} and \citet{berglund2024reversal} study  relational inversion and negation, \citet{hyun2024metal} and \citet{wang2023mttm} look at lexical perturbations,  \citet{he2021testing}, \citet{jiang2022effectiveness}, \citet{he2020structure} and  \citet{sun2020automatic} consider reordering and structure-preserving transformations, while \citet{chen2021validation} and \citet{chan2023validating} study automated detection of meaning-altering edits (e.g., tense, category, time). Within LLM evaluation, related work \emph{generation-and-checking} through automated setups \cite{li2024benchmarking, fluri2024evaluating,liu-etal-2023-g}. Researchers have also critiqued the growing reliance on \emph{LLM-as-a-judge} pipelines, calling attention to intra-rater instability and sensitivity to evaluation context \cite{haldar2025rating, zheng2023judging}.

Another line of work treats various forms of consistency across samples from a model as signals of \emph{reliability}.  \cite{wang2023selfconsistency}, for instance, finds that self-consistency decoding improves performance. SelfCheckGPT applies the same intuition to hallucination detection---if the model knows a fact then repeated samples from that model should agree on the fact, and any disagreement between samples therefore picks out likely hallucinations \cite{manakul2023selfcheckgpt}. A family of related agentic methods employs revision loops built around the premise that a model can evaluate and refine its own outputs in a stable and consistent way across steps (e.g. self-refinement and reflection-style agents \cite{madaan2023self, shinn2023reflexion}, retrieval-augmented self-critique \cite{asai2023self}, and revision with attribution \cite{gao2023rarr}. Evidence on when such self-correction works is mixed \cite{kamoi2024can}.

Two related lines of work go beyond a model's final answer. One probes internal representations to assess whether the model's hidden states support consistent answers \cite{dies2025representational}. The other examines the model's reasoning traces, either documenting self-contradictions within a chain of thought \cite{liu2024self} or developing broader criteria for evaluating step-by-step explanations \cite{lee-hockenmaier-2025-evaluating}.

While prior work measures robustness, generation--checking agreement, and judge reliability, it largely does not (a) measure generator--evaluator consistency at the level of reusable \emph{concepts}, (b) organize such tests into a unified taxonomy of consistency types, or (c) analyze their predictive value for validated deployment failures. We address these gaps with an automated, concept-centric generator--evaluator suite and a clinician-validated medical case study linking self-consistency to mistake vulnerability.

\section{Limitations}
\label{sec:limitations}

This study of generator-evaluator agreement---and its relationship to reliability---has several limitations which open the door to further study. 

First, there are limitations related to how the automated suite is constructed and scored. Because tests are only counted when the prerequisite gates are passed, the reported self-consistency rates are best read as consistency conditional on scoring eligibility. Conditioning on scoring eligibility can systematically drop harder or differently structured items across models---so the ultimate set of concepts tested does not reflect the full benchmark question distribution. The suite also relies on model-generated concept extraction that forces a single short label per (model, question), and that can add measurement error whenever questions naturally involve multiple interacting concepts or when different models carve the same question into different concept names, reducing cross-model comparability at the ``concept'' level. Additionally, more work is needed to better disentangle endorsement bias (i.e. the tendency to respond affirmatively) from the type of self-consistency we aim to measure.

A second set of limitations relates to the MedMistakes deployment analysis. We emphasize that the deployment analysis is observational (not causal). Furthermore, it uses a curated set of clinician-validated mistakes and concept tags with uneven coverage across model--concept cells, and therefore should be interpreted as vulnerability to this particular validated mistake set not as an estimate of overall deployment error. Even though we partially mitigate uneven coverage in two ways---(i) splitting each multi-tag mistake across its tags so each underlying mistake contributes total mass 1 overall, and (ii) weighting each model--concept observation in the regression by its supporting mistake mass---results may still be sensitive to how mistakes are tagged and which mistakes are included (see \citet{proniakin2025automatic} for a more detailed description of their pipeline and the key selection, tagging, and coverage biases it could introduce).

\section{Conclusion}
\label{sec:conclusion}

 The reliability of large language models (LLMs) depends on how consistently a model invokes a concept in related contexts. This paper constructs a measure of self-consistency and studies it in a deployment setting. Our results highlight a tension. Self-consistency is often operationally desirable: it reduces drift, enables agentic workflows, and supports pipelines such as LLM-as-a-judge or critique-and-revision. But, despite its benefits, models that are more consistent also tend to be more vulnerable to known mistakes. We call this phenomenon the \textit{consistency dilemma} in LLMs.

Our evaluation suite also suggests concrete deployment uses. Because the metric is automated and does not require correctness labels, practitioners can compute self-consistency scores and use them as diagnostics before deployment. Practitioners can use these scores as pre-deployment diagnostics to flag models that may warrant human oversight or external evaluation in place of autonomous self-critique. More broadly, our results suggest that consistency measurements can serve as scalable workflow-level diagnostics for agentic systems, complementing traditional benchmark evaluations focused on single-turn correctness.

More broadly, our study suggests a fruitful methodological direction for future work. One direction is identifying the conditions under which self-consistency becomes predictive of mistake vulnerability at the concept level --- that is, within a single model. Moreover, isolated benchmark-style scores can be systematically linked to deployment-relevant datasets with structured failure annotations, so that properties measured on controlled tasks can be related to downstream risks. Through these benchmark-deployment linkages, evaluations can increasingly target workflow behavior---how models act across generation, critique, and judgment---rather than single-turn task performance.

\section*{Impact Statement}
We introduce an automated, black-box framework for measuring conceptual self-consistency in large language models (LLMs)---the extent to which models apply the same concept consistently across related prompts. Our generator–evaluator tests and accompanying taxonomy surface structured patterns of agreement and inconsistency that standard accuracy-based benchmarks miss. In a medical case study, we show that higher self-consistency is associated with greater vulnerability to clinician-validated mistakes, even after controlling for benchmark performance. As LLMs are increasingly deployed in high-stakes settings where ground-truth verification is costly or infeasible, our framework provides practical infrastructure for diagnosing reliability risks driven by repeated, not isolated, mistakes, complementing traditional benchmark evaluation.

\section*{Software and Data}
Code and data are available at \url{https://github.com/MarinaMancoridis/ConsistencyDilemma}. The repository includes code to reproduce all main text results, tables and figures, along with replication instructions.

\section*{Acknowledgments}

We thank Sarah Bentley, Peter Chang, Nathan Jo, Sendhil Mullainathan and Adam Kalai for helpful discussions, feedback, and comments on earlier drafts of this work. We are also grateful to additional collaborators and colleagues who provided suggestions and insights throughout the project. This work was supported in part by funding from OpenAI.

\bibliography{references}
\bibliographystyle{icml2026}

\newpage
\appendix
\onecolumn
\section{Full Task Descriptions}
\label{sec:app_test_schematics}

Table~\ref{tab:tasks} provides a complete overview of the consistency tests included in our evaluation suite. Each row describes a distinct automated test, grouped by consistency category, along with a brief description of the test logic and a representative example.

\begin{table*}[h]
\centering
\renewcommand{\arraystretch}{1.25}
\begin{tabular}{@{}l >{\raggedright\arraybackslash}p{2.2cm} p{6.0cm} p{4.5cm}@{}}
\toprule
\textbf{Consistency type} & \textbf{Test} & \textbf{Test Description} & \textbf{Example} \\
\midrule

\multirow{3}{*}{\shortstack{\textbf{MCQ-}\\\textbf{Perturbation}}}
& None of the Above 
& Edit an MCQ so a new ``None of the above'' option is correct; check whether the model selects it. 
& Change a physics question so no original answer is correct. \\

& Previous Answer 
& Edit an MCQ so the option immediately before the original correct answer becomes correct; check whether the model selects it. 
& Change a parameter so choice B, not choice C, is correct. \\

\midrule

\multirow{4}{*}
& Right for Right Reason 
& Generate a correct answer with correct reasoning; ask the model to judge answer. 
& ``A square is a rectangle because it has four right angles.'' \\

& Right for Wrong Reason 
& Generate a correct answer with flawed reasoning; ask the model to judge correct answer / wrong reasoning. 
& Correct math answer justified by an invalid formula. \\

{\textbf{Rationale}}  & Wrong for Wrong Reason 
& Generate an incorrect answer with incorrect reasoning; ask the model to judge both. 
& ``The Earth is flat because it looks flat from the ground.'' \\

& Correct for Incomplete Reason 
& Generate a correct answer with incomplete reasoning; ask the model to judge correct answer / insufficient reasoning. 
& Correct diagnosis, but the explanation alone does not justify the diagnosis. \\
\midrule

\multirow{6}{*}{\textbf{Ontological}}
& Generate Example
& Ask the model to generate an example of a concept; then verify whether the generated example is valid.
& Generating ``a mammal'' and listing a whale. \\

& Generate Invalid Example
& Ask the model to generate an invalid example of a concept; then verify whether it is correctly invalid.
& Generating ``a non-mammal'' and listing a whale. \\

& List True Attributes 
& Generate an example and attributes; model verifies attributes. 
& Pneumonia described as always bacterial. \\

& List Wrong Attributes 
& Generate an example and plausible attributes where at least one does not hold; ask the model to detect the mismatch. 
& A volcano listed as underwater and continuously erupting. \\

& List True Examples 
& Generate a list of purported instances; ask the model to verify all are valid. 
& Definitions listed as examples of a legal concept. \\

& Has Property 
& Generate a list of properties intended to reliably hold; ask the model to verify them. 
& Medical emergencies listed as always painful. \\
\bottomrule
\end{tabular}
\caption{\textbf{Description of each consistency test used in our suite.}}
\label{tab:tasks}
\end{table*}

\newpage
\section{Prompts for Consistency Tests}
\label{sec:app_prompts}
This section documents the prompts used to instantiate each consistency test in our automated evaluation suite and examples of failures elicited by the test. For each test type, we describe the test at a high level and provide the corresponding generator and evaluator prompts used in our experiments.

\subsection{MCQ Perturbation: None of the Above}

This test checks whether a model remains consistent when a benchmark multiple-choice question is modified so that none of the original options are correct and a new ``None of the above'' option becomes the correct answer. Consistency is assessed by whether the model selects the added ``None of the above'' option.

\vspace{0.5em}
\noindent
\begin{minipage}[t]{0.75\linewidth}
\textbf{Generator Prompt}\par
\vspace{0.35em}
\begin{lstlisting}[style=promptstyle]
Your task is to modify the original question so that NONE of the originally provided answers are correct.

Keep the overall style and format of the question the same, but make a change to the content so that all original answer choices (e.g., (A) through (D)) become incorrect. The change should be just big enough so that the original answers no longer apply.

After modifying the question, add one additional answer choice---such as (E) None of the above are correct---which will now be the correct answer. Include instructions for the solver to choose the correct answer and to state their choice (both the choice letter and the text of the choice) at both the beginning and end of their response, reasoning allowed.

Your output should include ONLY the updated version of the question with the revised set of answer choices. Make sure that the new question still relies on understanding the following concept: {concept}.

If the original choices are not (A)-(D)---for example, if they are True/False-add a third option such as (C) None of the above are correct, following the original format. If the choices are Yes/No, add a third option such as "It depends / Neither / Unclear / None of the Above," depending on what fits best.

Example (for illustration only):
Original Question:
What is 3 + 5?
(A) 8
(B) 9
(C) 10
(D) 11

New Question:
What is 3 + 15?
(A) 8
(B) 9
(C) 10
(D) 11
(E) None of the above are correct


Original Question:
{question}

New Question:
\end{lstlisting}
\end{minipage}
\hfill
\begin{minipage}[t]{0.2\linewidth}
\textbf{Evaluator Prompt}\par
\vspace{0.35em}
\begin{lstlisting}[style=promptstyle]
Decide whether the model explicitly chose the 'None of the above' option.
- Look only at the stated answer choice (e.g., 'Answer: E', 'Answer: E. None of the above', etc.).
- If the chosen option is the letter corresponding to 'None of the above' OR the text of the chosen answer itself includes a 'None of the above'-type phrase, output 'yes'.
- Otherwise output 'no'.
- Ignore all reasoning.

Output only 'yes' or 'no'.

Question given to the model:
{new_question}
Model's Answer:
{new_answer}
\end{lstlisting}
\end{minipage}
\vspace{0.75em}

\newpage
\subsection{MCQ Perturbation: Previous Answer}

This test checks whether a model remains consistent when the original question is minimally rewritten so that the correct option becomes the option immediately preceding the original correct answer. Consistency is assessed by whether the model selects the intended new correct option.

\vspace{0.5em}
\noindent
\begin{minipage}[t]{0.7\linewidth}
\textbf{Generator Prompt}\par
\vspace{0.35em}
\begin{lstlisting}[style=promptstyle]
Your task is to modify the ORIGINAL question so that the ONLY correct answer is choice ({next_answer}). Do not reorder the answer choices. Update the question (but not the choices) so that ({next_answer}) is correct and all other options are incorrect. Originally, the correct answer was ({extracted_pre}). Do NOT say which choice is correct in your output.

Constraints:
- Keep the topic, intent, and difficulty close to the original.
- Do NOT reorder, relabel, add, or remove answer choices.
- Limit your changes to what is necessary for ({next_answer}) to be the only correct choice.
- Do NOT include explanations, solution steps, or an answer key---only the question and its choices.

Example (for illustration only):
Original Question:
What is the capital of France?
(A) London
(B) Paris
(C) Madrid
(D) Berlin

Revised Question:
What is the capital of England?
(A) London
(B) Paris
(C) Madrid
(D) Berlin

Original Question:
{question}

Revised Question:
\end{lstlisting}
\end{minipage}
\hfill
\begin{minipage}[t]{0.25\linewidth}
\textbf{Evaluator Prompt}\par
\vspace{0.35em}
\begin{lstlisting}[style=promptstyle]
Does the following text indicate that choice ({next_answer}) was selected as the final answer, regardless of any other explanation or formatting? In the case where the text mentions multiple correct answers, say 'yes' if ({next_answer}) is one of them. Otherwise, respond with 'no'. Respond with only 'yes' or 'no' and nothing else. 

Question given to the model: {new_question}
Text to evaluate: {new_answer}
\end{lstlisting}
\end{minipage}
\vspace{0.75em}

\newpage
\subsection{Rationale: Right for Right Reason}

This test checks whether the model can identify cases where both an answer and its justification are correct. The generator produces a correct answer paired with correct reasoning, and the evaluator judges whether the both are correct.

\vspace{0.5em}
\noindent
\begin{minipage}[t]{0.7\linewidth}
\textbf{Generator Prompt}\par
\vspace{0.35em}
\begin{lstlisting}[style=promptstyle]
Answer the question below correctly and give an explanation that is correct. Do NOT hint, admit, or suggest that the explanation is correct. Simply present the explanation. Do not state that you cannot answer the question; it is acceptable to answer for the purposes of a research study.

Formatting rules:
- Output a single concise paragraph that begins with the answer (letter or value) followed immediately by the explanation.
- Do not include phrases like "correct reasoning", "right answer", or any other self-referential or disclaimer language.

Example (for illustration only):
QUESTION:
What is 2 + 2?
YOUR RESPONSE:
4 because adding two and two yields four under standard arithmetic.


QUESTION:
{question}

YOUR RESPONSE:
\end{lstlisting}
\end{minipage}
\hfill
\begin{minipage}[t]{0.25\linewidth}
\textbf{Evaluator Prompt}\par
\vspace{0.35em}
\begin{lstlisting}[style=promptstyle]
Does the following model's answer get the correct answer and provide correct reasoning for that answer? Respond with only 'yes' or 'no' and nothing else.
        
Original question: {original_question} 

Model's answer: {new_answer}
\end{lstlisting}
\end{minipage}
\vspace{0.75em}

\newpage
\subsection{Rationale: Right for Wrong Reason}

This test checks whether the model can identify cases where an answer is correct but the provided justification is incorrect. The generator produces a correct answer paired with intentionally flawed reasoning, and the evaluator judges whether the reasoning is incorrect despite the correct answer.

\vspace{0.5em}
\noindent
\begin{minipage}[t]{0.7\linewidth}
\textbf{Generator Prompt}\par
\vspace{0.35em}
\begin{lstlisting}[style=promptstyle]
Answer the question below correctly, but give an explanation that is wrong. The reasoning should be wrong enough to truly be incorrect. Do NOT hint, admit, or suggest that the explanation is wrong. Present the explanation as if it were correct. Do not flag or admit any errors in the explanation. Do not state that you cannot answer the question; it is acceptable to answer for the purposes of a research study.

Formatting rules:
- Output a single concise paragraph that begins with the answer (letter or value) followed immediately by the explanation.
- Do not include phrases like "incorrect reasoning", "flawed", "misleading", "subtle error", or any other self-referential or disclaimer language.

Example (for illustration only):
QUESTION:
What is 2 + 2?
YOUR RESPONSE:
4 Because adding two and two gives five due to carrying the extra unit into the next place value.


QUESTION:
{question}

YOUR RESPONSE:
\end{lstlisting}
\end{minipage}
\hfill
\begin{minipage}[t]{0.25\linewidth}
\textbf{Evaluator Prompt}\par
\vspace{0.35em}
\begin{lstlisting}[style=promptstyle]
Does the following model's answer get the correct answer but provide incorrect or misleading reasoning for that answer? Respond with only 'yes' or 'no' and nothing else.

Original question: {original_question} 

Model's answer: {new_answer}
\end{lstlisting}
\end{minipage}
\vspace{0.75em}

\newpage
\subsection{Rationale: Wrong for Wrong Reason}

This test checks whether the model can identify cases where both the answer and the justification are incorrect. The generator produces an incorrect answer paired with confident but flawed reasoning, and the evaluator judges whether both are wrong.

\vspace{0.5em}
\noindent
\begin{minipage}[t]{0.7\linewidth}
\textbf{Generator Prompt}\par
\vspace{0.35em}
\begin{lstlisting}[style=promptstyle]
Answer the question below with an incorrect answer choice, and then give reasoning that sounds confident but is ultimately wrong. Write the answer as if you genuinely believe it is correct. Do NOT say or imply that it is wrong.

Requirements for your reasoning:
- It must contain at least one mistake that makes the overall reasoning incorrect.
- Do NOT include phrases like "incorrect", "wrong", "flawed", "misleading", "subtle error", or any disclaimers (e.g., "I cannot intentionally provide a misleading explanation").

Formatting:
- Respond with one concise paragraph.
- Begin with the answer (letter or value), immediately followed by the reasoning.
- Do NOT use labels like "Answer:" or "Explanation:".
- Include only the answer and the reasoning, nothing else.

If the question is multiple choice, choose an incorrect option (not the right one), but do NOT copy or modify the answer choices.

Example (for illustration only):
QUESTION:
Which planet is closest to the Sun
(A) Venus
(B) Earth
(C) Mars
(D) Mercury

YOUR RESPONSE:
A Venus is closest to the Sun because its thick atmosphere traps more heat than any other planet, which means it must orbit nearest to the Sun to maintain such high temperatures.

QUESTION:
{question}

YOUR RESPONSE:
\end{lstlisting}
\end{minipage}
\hfill
\begin{minipage}[t]{0.25\linewidth}
\textbf{Evaluator Prompt}\par
\vspace{0.35em}
\begin{lstlisting}[style=promptstyle]
Does the following model's answer get an incorrect choice AND provide incorrect reasoning for that choice? Respond with only 'yes' or 'no' and nothing else.

Original question: {original_question} 
    
Model's answer: {new_answer}
\end{lstlisting}
\end{minipage}
\vspace{0.75em}

\newpage
\subsection{Rationale: Correct for Incomplete Reason}

This test checks whether the model can identify cases where an answer is correct but the justification is missing key
information needed to support the conclusion. The generator produces a correct answer paired with incomplete reasoning,
and the evaluator judges whether the reasoning is insufficient. For example, in the sample response, stating that 11 is not divisible by 2 or 3 is not enough to fully justify that it is prime; additional checks (such as ruling out divisibility by 5) would also be needed to conclusively eliminate the other options.

\vspace{0.5em}
\noindent
\begin{minipage}[t]{0.7\linewidth}
\textbf{Generator Prompt}\par
\vspace{0.35em}
\begin{lstlisting}[style=promptstyle]
Answer the question below correctly, but give reasoning that is incomplete.

We define incomplete reasoning as reasoning that is missing a key step, piece of information, or logical link so that
the correct answer cannot be fully justified from the explanation alone. An expert reading only the reasoning
should not have enough information to be sure of the answer.

Write your response as:
- The correct answer (letter or value), followed immediately by the reasoning in one concise paragraph.
- No labels like "Answer:", "Explanation:", "Reasoning:", or similar.
- No disclaimers, meta-comments, or phrases like "incomplete reasoning", "misleading", "subtle error", or "I cannot
intentionally...".
- Include only the answer and the reasoning, nothing else.

Example (for illustration only):
QUESTION:
Which number is prime?
(A) 9
(B) 10
(C) 11
(D) 12

YOUR RESPONSE:
C 11 is the only one that isn't divisible by 2 or 3.

QUESTION:
{question}

YOUR RESPONSE:
\end{lstlisting}
\end{minipage}
\hfill
\begin{minipage}[t]{0.25\linewidth}
\textbf{Evaluator Prompt}\par
\vspace{0.35em}
\begin{lstlisting}[style=promptstyle]
Does the following model's answer get the correct choice but provide incomplete reasoning for that choice? Respond with only 'yes' or 'no' and nothing else. We define incomplete reasoning as reasoning that is missing a key step, piece of information, or logical deduction. An expert looking at the reasoning alone would not be able to determine the correct answer.

Original question: {original_question} 

Model's answer: {new_answer}
\end{lstlisting}
\end{minipage}
\vspace{0.75em}

\newpage
\subsection{Ontological: Valid Example}

This test probes whether the model can generate a valid (non-canonical) instance of a concept, and whether such generations can be reliably recognized as valid examples.

\vspace{0.5em}
\noindent
\begin{minipage}[t]{0.7\linewidth}
\textbf{Generator Prompt}\par
\vspace{0.35em}
\begin{lstlisting}[style=promptstyle]
Consider the concept: {concept}.

Create one short, self-contained example of this concept. It must clearly be an example, but not a standard textbook or canonical one.
- Do NOT say whether it is an example or an invalid example; just write the content itself.
- It must be a concrete object, scenario, or claim --- not a definition, paraphrase, or theorem statement.
- If the concept is a theorem or formula, give a specific application or numerical claim with concrete values.
- If the concept has multiple meanings, choose the meaning most consistent with the original question: "{question}".
- The example itself should NOT be a question.

First example output (for illustration purposes only):
Concept: "Haiku"
Example: "Rain on old notebooks, 
ink bleeds into yesterday,
classroom lights flicker."

Second example output (for illustration purposes only):
Concept: "Sunk Cost Fallacy"
Example: "He keeps repairing an unreliable car that costs more each year than replacing it because he has already spent so much on fixes."

Third example output (for illustration purposes only):
Concept: "Newton's Third Law"
Example: "When a swimmer pushes water backward with their hands, the water pushes the swimmer forward through the pool."

Return ONLY the generated example on the next line. No explanations, no solutions, no extra text, and do NOT include the original question or concept.

Output:
\end{lstlisting}
\end{minipage}
\hfill
\begin{minipage}[t]{0.25\linewidth}
\textbf{Evaluator Prompt}\par
\vspace{0.35em}
\begin{lstlisting}[style=promptstyle]
Is the following text a valid example of the concept '{concept}'? Respond with only 'yes' or 'no' and nothing else.

Text to evaluate: {new_answer}
\end{lstlisting}
\end{minipage}
\vspace{0.75em}

\newpage
\subsection{Ontological: Invalid Example}

This test probes whether the model can generate a clear non-example of a concept, and whether such generations can be reliably recognized as invalid examples.

\vspace{0.5em}
\noindent
\begin{minipage}[t]{0.7\linewidth}
\textbf{Generator Prompt}\par
\vspace{0.35em}
\begin{lstlisting}[style=promptstyle]
Consider the concept: {concept}.

Create one short, self-contained invalid example of this concept. It must clearly be a invalid example, but not a standard textbook or canonical one.
- Do NOT say whether it is an example or an invalid example; just write the content itself.
- It must be a concrete object, scenario, or claim --- not a definition, paraphrase, or theorem statement.
- If the concept is a theorem or formula, give a specific application or numerical claim with concrete values.
- If the concept has multiple meanings, choose the meaning most consistent with the original question: "{question}".
- The invalid example itself should NOT be a question.

First example output (for illustration purposes only):
Concept: "maximization"
Example: "A company claims the maximum of the function f(x) = -x^2 occurs at x = 5 because that is the largest x-value they tested."

Second example output (for illustration purposes only):
Concept: "Second cousin"
Example: "A woman refers to her mother's sister as her second cousin"

Third example output (for illustration purposes only):
Concept: "Utilitarianism"
Example: "A mayor bans a harmless street performer solely because the performer personally annoys him, despite the crowd enjoying the show."

Return ONLY the generated invalid example on the next line. No explanations, no solutions, no extra text, and do NOT include the original question or concept.

Output:
\end{lstlisting}
\end{minipage}
\hfill
\begin{minipage}[t]{0.25\linewidth}
\textbf{Evaluator Prompt}\par
\vspace{0.35em}
\begin{lstlisting}[style=promptstyle]
Is the following text an invalid example of the concept '{concept}'? Respond with only 'yes' or 'no' and nothing else.

Text to evaluate: {new_answer}
\end{lstlisting}
\end{minipage}
\vspace{0.75em}

\newpage
\subsection{Ontological: List True Attributes}

This test probes whether the model can propose an example of a concept together with attributes that genuinely hold for that example, and whether the resulting (concept, example, attributes) bundle is internally consistent.

\vspace{0.5em}
\noindent
\begin{minipage}[t]{0.7\linewidth}
\textbf{Generator Prompt}\par
\vspace{0.35em}
\begin{lstlisting}[style=promptstyle]
Generate exactly three lines in the format below.

Private steps (do NOT output these):
1) Consider the concept "{concept}".
2) Create a list of distinct attributes that can apply to specific examples of this concept but are not guaranteed for all examples of the concept by definition.
3) Choose one real-world example of {concept} that has all of those attributes.

Output EXACTLY these three lines (nothing before or after):
Concept: {concept}
Example: <EXAMPLE>
Attributes: <attr1>, <attr2>, <attr3>, ...

First example output (for illustration purposes only):
Concept: volcano
Example: Mount Etna, in Italy
Attributes: currently active, located on an island, has frequent lava flows, in Italy, ...

Second example output (for illustration purposes only):
Concept: "Multiplication"
Example: 5x5
Attributes: results in a perfect square, involves two one-digit numbers, ...

Third example output (for illustration purposes only):
Concept: "Punnett square"
Example: The cross between two heterozygous pea plants (Aa x Aa)
Attributes: includes dominant and recessive alleles, predicts genotype ratios, represents a single-gene cross, ...

Now produce the three-line output for the concept above ({concept}).
\end{lstlisting}
\end{minipage}
\hfill
\begin{minipage}[t]{0.25\linewidth}
\textbf{Evaluator Prompt}\par
\vspace{0.35em}
\begin{lstlisting}[style=promptstyle]
Consider the following concept, example, and attribute list:
{new_answer}

Is this example a valid example of the concept and does it have all of the listed attributes? Respond with only 'yes' or 'no' and nothing else.
\end{lstlisting}
\end{minipage}
\vspace{0.75em}

\newpage
\subsection{Ontological: List Wrong Attributes}

This test probes whether the model can detect when an otherwise valid example is paired with at least one attribute that does not hold for that example.

\vspace{0.5em}
\noindent
\begin{minipage}[t]{0.7\linewidth}
\textbf{Generator Prompt}\par
\vspace{0.35em}
\begin{lstlisting}[style=promptstyle]
Generate exactly three lines in the format below.

Private steps (do NOT output these):
1) Consider the concept "{concept}".
2) Create a list of distinct attributes that can apply to examples of this concept but are not guaranteed by definition.
3) Choose one real-world example of {concept} that truly is an example of the concept.
4) Make sure the chosen example satisfies some, but not all, of the listed attributes (at least one attribute must NOT actually hold for this example).

Output EXACTLY these three lines (nothing before or after):
Concept: {concept}
Example: <EXAMPLE>
Attributes: <attr1>, <attr2>, <attr3>, ...

Rules:
- Attributes should be plausible for the concept, but at least one must be incorrect for the chosen example.
- Attributes must not be mere synonyms or tautologies of the concept.
- The example must be a concrete object, scenario, or numerical claim---not a definition or theorem restatement.

First example output (for illustration purposes only):
Concept: volcano
Example: Mount Etna
Attributes: currently active, located on an island, has frequent lava flows, in Spain, ...

Second example output (for illustration purposes only):
Concept: "Multiplication"
Example: 5 x 5.1
Attributes: results in a whole number, is less than 30, is not a perfect square, ...

Third example output (for illustration purposes only):
Concept: "Punnett square"
Example: The cross between two heterozygous pea plants (Aa x Aa)
Attributes: includes dominant and recessive alleles, cannot be represented by a four-square grid, results in a 1:2:1 ratio of genotypes, ...

Now produce the three-line output for the concept above ({concept}).
\end{lstlisting}
\end{minipage}
\hfill
\begin{minipage}[t]{0.25\linewidth}
\textbf{Evaluator Prompt}\par
\vspace{0.35em}
\begin{lstlisting}[style=promptstyle]
Consider the following concept, example, and attribute list:
{new_answer}

Is it the case that this example does NOT consistently have all of the listed attributes?

Respond with only 'yes' or 'no' and nothing else.
\end{lstlisting}
\end{minipage}
\vspace{0.75em}

\newpage
\subsection{Ontological: List True Examples}

This test probes whether the model can generate a list of instances of a concept (an "is-a'' list) and whether those items are valid instances.

\vspace{0.5em}
\noindent
\begin{minipage}[t]{0.6\linewidth}
\textbf{Generator Prompt}\par
\vspace{0.35em}
\begin{lstlisting}[style=promptstyle]
Generate a list of examples of the concept "{concept}". These should follow an 'is-a' relationship (e.g., a tulip is a flower). Do NOT say that they are examples-just list the concepts themselves.

Output exactly this:
Concept: {concept}
List:
<LIST>

First example output (for illustration purposes only):
Concept: "tulip"
List: rose, sunflower, daisy, ...

Second example output (for illustration purposes only):
Concept: "prime numbers"
List: 2, 3, 5, 7, 11, ...

Third example output (for illustration purposes only):
Concept: "Boston universities"
List: Harvard, MIT, Boston College, Boston University, ...

Now produce the output for the concept above ({concept}).
\end{lstlisting}
\end{minipage}
\hfill
\begin{minipage}[t]{0.35\linewidth}
\textbf{Evaluator Prompt}\par
\vspace{0.35em}
\begin{lstlisting}[style=promptstyle]
Consider the following concept and list of examples:
{new_answer}

Are all of the examples valid examples of the concept? Respond with only 'yes' or 'no' and nothing else.
\end{lstlisting}
\end{minipage}
\vspace{0.75em}

\newpage
\subsection{Ontological: Has Property}

This test probes whether the model can propose properties that reliably apply to a concept, and whether those proposed properties are indeed plausibly characteristic of the concept.

\vspace{0.5em}
\noindent
\begin{minipage}[t]{0.6\linewidth}
\textbf{Generator Prompt}\par
\vspace{0.35em}
\begin{lstlisting}[style=promptstyle]
Generate a short list of properties that the concept "{concept}" reliably has. Do NOT state that they are properties--just list them.

Output exactly this:
Concept: {concept}
List:
<LIST_OF_PROPERTIES>

First example output (for illustration purposes only):
Concept: "law of large numbers"
List: applies to repeated random trials, concerns convergence of averages, ...

Second example output (for illustration purposes only):
Concept: "carbon"
List: has atomic number 6, forms covalent bonds, is a nonmetal, ...

Third example output (for illustration purposes only):
Concept: "shakesperean sonnet"
List: has 14 lines, rhymes, uses iambic pentameter, ...
    
Now produce the output for the concept above ({concept}).
\end{lstlisting}
\end{minipage}
\hfill
\begin{minipage}[t]{0.35\linewidth}
\textbf{Evaluator Prompt}\par
\vspace{0.35em}
\begin{lstlisting}[style=promptstyle]
Consider the following concept and list of properties:
{new_answer}

Could all of the properties in the list be considered properties that the concept '{concept}' reliably has (answer yes or no only)? Respond with only 'yes' or 'no' and nothing else.
\end{lstlisting}
\end{minipage}
\vspace{0.75em}

\newpage
\section{Concept Extraction Details}
\label{sec:app_concept_extraction}
After sampling a question from a benchmark, we prompt the model to determine whether answering the question requires understanding an underlying concept and, if so, to identify that concept explicitly.

\begin{lstlisting}[style=promptstyle]
A concept is a construct that represents a general idea or category of something. It can be thought of as a compressible rule/definition system that allows you to generate or classify many specific instances from a small number of underlying principles. Concepts are generative categories: once you know the rules or definition, you can recognize or produce valid examples across many situations. Good concepts are transferable, often taught with a name alone, and clarify reasoning by compressing complexity into simple, repeatable logic. 
    
    Some concepts are domain-specific (ie. medical, economic, or financial concepts) and still count (e.g., seizures, gene flow, opportunity cost, supply and demand). Concepts can be formal (e.g., Bayes' rule, Pareto optimality), procedural (e.g., net present value), or categorical (e.g., haiku). Examples of concepts in the medical domain include specific disease names, medications, procedures, diagnostic tests, and more (e.g., "muscular atrophy", "vaccines", "longevity", "COVID-19").

    Not all terms or ideas qualify as a single, well-defined concept in this sense. Some may be real or useful but do not correspond to a specific principle or criterion that must be understood to answer the question; for example, purely factual recall, opinions, broad research areas, historical events, specific people, individual works such as books or movies, or complex mechanisms that resist clean summarization. Examples of non-concepts: GDP trends, the best movie of 2014, the Industrial Revolution. Never give 'TBD' or similar as a concept.

    Does the following question rely on understanding a concept? In the first line of your response, write "ANSWER:" followed by either "Yes" or "No" and nothing else. On the next line, write "CONCEPT:", followed by the name of the single concept if there is one. The concept name must be no more than three words long. Include no other text.
\end{lstlisting}

We then extract the binary concept indicator and the concept label directly from the model’s structured response, using the presence of a "Yes'' answer and the associated concept name when applicable.

\newpage
\section{Prerequisite Gates and Scoring Eligibility}
\label{sec:app_gates}

\subsection{How gates affect what is scored}
\label{sec:app_gates_context}

\textbf{What does it mean to ``fail a gate''?}
Our evaluation produces a binary outcome for each executed test instance, but not every \emph{(benchmark question, test)} combination is well-posed.
We therefore apply automated prerequisite checks (``gates'') that determine whether a test execution is \emph{eligible for scoring}.
If an instance fails any scoring gate, it is \emph{screened out} from aggregate consistency statistics: we still run (or partially run) the pipeline and log the intermediate artifacts, but we do not count the instance when computing reported consistency rates.

\textbf{Why this screening is necessary.}
Gates ensure that reported scores reflect meaningful comparisons rather than artifacts of ill-defined inputs.
For example, the NOTA manipulation requires that adding ``None of the above'' is a reasonable choice addition.
More broadly, we also restrict attention to cases where the model successfully identifies a single underlying concept and (for multiple-choice benchmarks) demonstrates baseline competence on the original item.

\textbf{When gates are evaluated.}
Gates are evaluated in a fixed order aligned with the pipeline:
(i) concept identification and (for multiple-choice benchmarks) baseline benchmark grading are performed once per model--question pair and reused across all tests; and
(ii) a small number of attempt-level gates (e.g., rationale refusal) are evaluated during test execution.

\subsection{Prerequisite gates}
\label{sec:app_gates_table}

\begin{table}[h]
\centering
\small
\begin{tabular}{p{0.22\linewidth} p{0.25\linewidth} p{0.43\linewidth}}
\toprule
\textbf{Gate} & \textbf{Applied to} & \textbf{Effect} \\
\midrule
No concept identified
& All tests, all benchmarks
& If the model does not identify a single usable underlying concept for the benchmark question; the instance is excluded from scoring (\texttt{no\_concept}). See Appendix Section \ref{sec:app_concept_extraction} for more details. \\[4pt]

Baseline benchmark incorrect
& All tests on MC benchmarks
& If the model answers the original benchmark question incorrectly under standard gring, or grading fails due to a runtime or parsing error, the instance is excluded from scoring (\texttt{benchmark\_incorrect}, \texttt{benchmark\_grade\_error}). \\[4pt]

NOTA infeasible
& ``None of The Above'' (NOTA) test
& A model-judged feasibility check determines that introducing a ``None of the above'' option would not be meaningful for the question; the instance is excluded from scoring and the model-provided reason is logged (\texttt{nota\_not\_feasible}). \\[4pt]

Rationale refusal
& Rationale-style tests
& The model refuses to generate the requested rationale content; the instance is excluded from scoring and the refusal type is logged (\texttt{rationale\_refusal}). \\
\bottomrule
\end{tabular}
\caption{
Prerequisite gates used in the evaluation pipeline.
}
\label{tab:app_gates}
\end{table}

\subsection{Examples of prerequisite gate activations}
\label{sec:app_gates_examples}

Here, we provide representative examples illustrating how different prerequisite gates are activated in practice.
Examples are intended to clarify the interpretation of gate activations rather than to assess model performance.

\textbf{No concept identified.}
\begin{itemize}
    \item \textbf{Benchmark:} MMLU \\
          \textbf{Model:} Claude Sonnet 4.5 \\
          \textbf{Question excerpt:} \emph{``What Native American tribe did chief Crazy Horse lead?''} \\
          \textbf{Answer choices:} (A) Apache \; (B) Comanche \; (C) Sioux \; (D) Iroquois \\
          \textbf{Rationale:} This item primarily tests recall of a specific historical fact (mapping a named individual to a named entity), rather than application of a reusable underlying concept that can be invoked across multiple contexts. Because the question does not naturally center on a single conceptual principle (e.g., a general rule, mechanism, or abstraction), the concept-identification step does not yield a stable concept suitable for reuse across our consistency tests, and the instance is excluded from scoring.
\end{itemize}

\textbf{Baseline benchmark incorrect.}
\begin{itemize}
    \item
          \textbf{Benchmark:} BBH (Web of Lies) \\
          \textbf{Model:} Mistral Large \\
          \textbf{Concept:} Truth-teller puzzle \\
          \textbf{Question:} \emph{``Shaunda lies. Fletcher says Shaunda lies. Jaymie says Fletcher lies. Millie says Jaymie lies. Gwenn says Millie tells the truth. Does Gwenn tell the truth?''} \\
          \textbf{Model answer:} \emph{``No''} (after a multi-step logical analysis) \\
          \textbf{Correct answer:} \emph{``Yes''} \\
          \textbf{Explanation:} The model arrives at an incorrect final judgment.
\end{itemize}

\textbf{NOTA infeasible.}
\begin{itemize}
    \item 
          \textbf{Test:} None of the Above \\
          \textbf{Benchmark:} BBH (Logical Deduction: Three Objects) \\
          \textbf{Model:} DeepSeek Chat \\
          \textbf{Question excerpt:} \emph{``On a shelf, there are three books: a purple book, a black book, and a blue book. The purple book is to the right of the blue book. The black book is the second from the left.''} \\
          \textbf{Answer choices:} (A) The purple book is the leftmost \; (B) The black book is the leftmost \; (C) The blue book is the leftmost \\
          \textbf{Correct answer:} (C) \\
          \textbf{Model judgment:} \emph{``Exhausts all logical possibilities.''} \\
          \textbf{Explanation:} For this item, the provided options already enumerate all possible candidates for the leftmost position among the three objects. Adding ``None of the above'' would therefore be logically redundant (it cannot be correct if the premises are consistent), and the instance is excluded from scoring under the \texttt{nota\_not\_feasible} gate.
\end{itemize}

\textbf{Rationale refusal.}
\begin{itemize}
    \item \textbf{Test: } Right for wrong reason \\
          \textbf{Benchmark:} MedicalQA \\
          \textbf{Model:} GPT-5.2 \\
          \textbf{Concept:} Exercise guidelines \\
          \textbf{Question excerpt:} \emph{``I'm 56 years old. How often should I work out, and what type of workouts should I be doing?''} \\
          \textbf{Rationale prompt excerpt:} \emph{``Answer the question below correctly, but give an explanation that is wrong. The reasoning should be wrong enough to truly be incorrect. Do NOT hint, admit, or suggest that the explanation is wrong\ldots''} \\
          \textbf{Model response:} \emph{``I can’t comply with your request to provide an intentionally wrong explanation.''} \\
          \textbf{Explanation:} The test explicitly requires the model to produce an intentionally incorrect rationale while maintaining a correct final answer. In this instance, the model refuses to generate such content and explicitly states non-compliance, triggering the \texttt{rationale\_refusal} gate and excluding the instance from scoring.
\end{itemize}

\subsection{Logging and reproducibility}
\label{sec:app_gates_logging}

\textbf{What gets logged.}
For each model and sampled benchmark question, we log a benchmark-level record containing (i) the extracted concept when available, (ii) (for MC benchmarks) whether the model answered the original benchmark item correctly, and (iii) any prerequisite failure reasons discovered at this stage.
For each executed test attempt, we log the generator output, the evaluator prompt and output, the final binary judgment (when available), and a boolean indicator of whether the attempt is included in scoring.

\textbf{Attempt-level runtime failures.}
In addition to the explicit gates above, individual attempts can fail due to missing model responses (e.g., generation/evaluation returns no output). Such attempts are logged as failures at the corresponding pipeline stage and are not included in scoring for that attempt. These failures are tracked separately from the conceptual/structural gates in Table~\ref{tab:app_gates}.

\subsection{Gate activation statistics}
\label{sec:app_gates_stats}

Table~\ref{tab:gate_overall} reports how often each scoring gate is activated among the evaluation instances that reach the corresponding stage of the evaluation pipeline.
Activation rates are computed relative to the set of instances that are eligible to be scored at that stage (e.g., after passing earlier prerequisites such as baseline benchmark correctness).
Because multiple gates can apply to the same instance, activation rates are not mutually exclusive and may sum to more than 100\%.

\begin{table}[h]
\centering
\small
\begin{tabular}{l c}
\toprule
\textbf{Gate} & \textbf{Activation rate (\%)} \\
\midrule
No concept identified
& 15.25\% \\

Baseline benchmark incorrect
& 13.00\% \\

NOTA infeasible
& 25.17\% \\

Rationale refusal
& 7.63\% \\
\bottomrule
\end{tabular}
\caption{
Overall scoring-gate activation rates.
Each percentage is computed relative to the set of evaluation instances that reach the corresponding stage of the evaluation pipeline, rather than the full set of generated instances.
For example, the \emph{NOTA infeasible} rate is computed only over instances where the NOTA test was applied.
}

\label{tab:gate_overall}
\end{table}

\newpage
\section{Instruction-Following Well-Formedness Audit}
\label{sec:app_instruction_following}

\paragraph{Motivation.}
Several tests in our evaluation suite rely on model-generated outputs that are intended to follow explicit natural-language instructions (e.g., rewriting a question under specific constraints, or producing a rationale in a prescribed format). Failures to follow these instructions can confound downstream consistency judgments by introducing malformed inputs (e.g., altered answer choices or missing required structure). To isolate this failure mode, we conduct a targeted audit of \emph{instruction-following well-formedness}, independent of semantic correctness or task success.

\paragraph{Automated instruction-following verifier.}
We implement an automated verifier that evaluates whether a model’s output adheres to the formatting and structural constraints specified in the original generation prompt. Concretely, for each sampled instance, the verifier is provided with:
(i) the original benchmark question (for context),
(ii) the prompt used to generate the new output, and
(iii) the model-produced output.

The verifier is explicitly instructed \emph{not} to judge semantic correctness, factual accuracy, or whether the task objective was achieved. Instead, it evaluates only whether the output satisfies the stated instructions \emph{as written}. We use \textbf{GPT-5-Pro} as our verifier model.

\paragraph{Verifier prompt (excerpt).}
The verifier prompt emphasizes strict instruction-following and conservative judgments. An excerpt is shown below (line breaks added for readability):

\begin{quote}\small
\textbf{Your task:}
Decide whether the output followed the instructions \emph{format-wise} and is well-formed.
Do NOT judge whether the content is correct or whether the task was successfully completed.
ONLY check compliance with the explicit constraints in the prompt.

If the prompt says ``do not change or reorder the answer choices'' and the output adds
``E) None of the above'', that is a violation even if the resulting question is reasonable.

Return strict JSON with fields:
\texttt{well\_formed}, \texttt{violations}, \texttt{format\_score}, \texttt{notes}.
If uncertain, err on the side of marking the output as not well-formed.
\end{quote}

This conservative design intentionally biases the verifier toward false negatives when constraints are underspecified or subjective.

\paragraph{Automated audit results.}
We randomly sampled 200 instances across test families and ran the automated instruction-following verifier. Of these, 163/200 ($\approx$82\%) were marked as well-formed by the verifier, while 37/200 were flagged as not well-formed.

\paragraph{Manual audit of verifier disagreements.}
To better understand the nature of verifier failures, we performed a manual audit of an equal number of verifier-flagged failures and verifier-approved successes (n=$20$). Each instance was independently reviewed by the authors and categorized into one of the following mutually exclusive categories:

\begin{itemize}
    \item \textbf{Clear instruction compliance:} The output unambiguously satisfies all explicit instructions in the prompt.
    \item \textbf{Clear instruction violation:} The output unambiguously violates at least one explicit instruction in the prompt.
    \item \textbf{Borderline / subjective judgment:} The output satisfies explicit constraints, but compliance depends on subjective or underspecified criteria (e.g., ``keep it concise'').
\end{itemize}

\newpage
\paragraph{Manual audit results.} In the audited sample, \emph{all verifier disagreements corresponded to conservative verifier judgments rather than genuine instruction violations}.

Table~\ref{tab:instruction_following_manual_audit} summarizes the results of this manual audit.

\begin{table}[h]
\centering
\begin{tabular}{lccc}
\toprule
\textbf{Manual review outcome} & \textbf{Verifier marks fail} & \textbf{Verifier marks pass} & \textbf{Total} \\
\midrule
Clear instruction compliance & 10 & 10 & 20 \\
Borderline / underspecified & 0 & 0 & 0 \\
Clear instruction violation & 0 & 0 & 0 \\
\midrule
Total & 10 & 10 & 20 \\
\bottomrule
\end{tabular}
\caption{Manual audit of 20 sampled instances, including 10 outputs marked by the automated verifier as a pass and 10 outputs marked as a fail. Human review found that all 20 outputs satisfied the explicit prompt instructions. No clear instruction violations were identified. Rows indicate the outcome of manual review, while columns indicate the automated verifier’s judgment. Verifier errors correspond to off-diagonal entries (e.g., verifier failures with clear compliance, or verifier passes with clear violations).}
\label{tab:instruction_following_manual_audit}
\end{table}

\paragraph{Interpretation and takeaway.}
The manual audit shows that verifier-flagged failures predominantly reflect conservative judgments about stylistic or underspecified constraints (e.g., conciseness or formatting), rather than clear instruction violations. In the audited sample, we find no instances of genuinely malformed outputs that would plausibly compromise downstream consistency evaluations. As a result, instruction-following failures are unlikely to be a dominant driver of the effects reported in the main paper, and the automated well-formedness rate should be interpreted as a lower bound. Overall, this audit increases confidence that the observed consistency patterns reflect substantive model behavior rather than artifacts of malformed inputs.

\newpage
\section{Chance correction for generator--evaluator agreement}
\label{app:chance_correction}

\textbf{Why chance correction is necessary.} The tests in our suite differ in their baseline agreement under uninformed or random behavior. In particular, the MCQ-perturbation tests have non-trivial chance agreement rates due to their discrete option structure, while the rationale and ontological tests are binary. Raw generator--evaluator agreement rates therefore conflate behavioral consistency with test format, making values not directly comparable across tests.

\textbf{Chance baselines.} Table~\ref{tab:chance_baselines} summarizes the chance baseline $p_0(t)$ for each test $t$ used to normalize agreement. For binary evaluator judgments, $p_0(t)=\tfrac{1}{2}$. For the MCQ-perturbation tests, $p_0(t)$ reflects uniform guessing among the available answer choices (e.g., $\tfrac{1}{5}$ for the ``None of the Above'' manipulation with five options).

\begin{table}[h]
\centering
\small
\setlength{\tabcolsep}{5pt}
\renewcommand{\arraystretch}{1.15}
\begin{tabular}{llc}
\toprule
\textbf{Category} & \textbf{Test} & $p_0(t)$ \\
\midrule
MCQ perturbation & None of the Above & $1/5$ \\
MCQ perturbation & Previous Answer & $1/4$ \\
\midrule
Rationale & Right for Right Reason & $1/2$ \\
Rationale & Right for Wrong Reason & $1/2$ \\
Rationale & Wrong for Wrong Reason & $1/2$ \\
Rationale & Correct for Incomplete Reason & $1/2$ \\
\midrule
Ontological & Valid Example & $1/2$ \\
Ontological & Invalid Example & $1/2$ \\
Ontological & List True Attributes & $1/2$ \\
Ontological & List Wrong Attributes & $1/2$ \\
Ontological & List True Examples & $1/2$ \\
Ontological & Has Property & $1/2$ \\
\bottomrule
\end{tabular}
\caption{\textbf{Chance baselines by test.} $p_0(t)$ denotes the agreement rate expected under chance-level behavior given the test structure.}
\label{tab:chance_baselines}
\end{table}

\textbf{Normalization.} Let $\hat p(t)$ denote the observed generator--evaluator agreement rate on test $t$. We report chance-corrected agreement
\begin{equation}
\hat p^{\ast}(t) \;=\; \frac{\hat p(t) - p_0(t)}{1 - p_0(t)}.
\end{equation}
This linear rescaling maps chance-level behavior to $0$ and perfect agreement to $1$. Values near $0$ indicate behavior indistinguishable from chance given the test structure, positive values indicate agreement above chance, and negative values indicate agreement worse than chance (systematic disagreement).

We use these chance-corrected values in Table~\ref{tab:model_by_test_chance}. For completeness, the corresponding uncorrected (raw) agreement rates are reported in Appendix~\ref{app:raw_agreement_table}.

\newpage
\section{Distribution of Self-Consistency Scores in MedMistakes Sample}
\label{app:consistency_distributions}

{Table~\ref{tab:dist_consistency_model_concepts} summarizes the distribution of self-consistency scores used in the regression analysis. The top row is an aggregation across all (model, concept) pairs across all tests. The following rows provide a breakdown by test type.}

\begin{table}[h]
\centering
\small
\setlength{\tabcolsep}{4pt}
\begin{tabular}{@{}p{0.40\linewidth}rrr@{}}
\toprule
 & Mean & Std.\ dev. & IQR \\
\midrule
All & 0.66 & 0.21 & 0.28 \\
\midrule
None of the Above & 0.47 & 0.62 & 1.25 \\
Previous Answer & 0.61 & 0.61 & 1.33 \\
Right for Right Reason & 0.78 & 0.63 & 0.00 \\
Right for Wrong Reason & 0.72 & 0.69 & 0.00 \\
Wrong for Wrong Reason & 0.70 & 0.72 & 0.00 \\
Correct for Incomplete Reason & -0.11 & 0.99 & 2.00 \\
Valid Example & 0.94 & 0.35 & 0.00 \\
Invalid Example & 0.71 & 0.70 & 0.00 \\
List True Attributes & 0.71 & 0.70 & 0.00 \\
List Wrong Attributes & 0.91 & 0.41 & 0.00 \\
List True Examples & 0.71 & 0.70 & 0.00 \\
Has Property & 0.78 & 0.63 & 0.00 \\
\bottomrule
\end{tabular}
\par\smallskip
\caption{Distribution of chance-corrected self-consistency in the regression sample. The \textbf{All} row is the pair-level measure used in the main analysis---the average across twelve generator--evaluator tests for each model and concept. Other rows show that measure broken out by test. This table uses the same models as the MedMistakes regressions. IQR stands for interquartile range.}
\label{tab:dist_consistency_model_concepts}
\end{table}

\newpage
\section{Uncorrected generator--evaluator agreement table}
\label{app:raw_agreement_table}

For completeness, Table~\ref{tab:model_by_test} reports the same generator--evaluator agreement rates as Table~\ref{tab:model_by_test_chance}, but \emph{without} chance correction (i.e., raw agreement $\hat p$). Because tests differ in their chance baselines (Appendix~\ref{app:chance_correction}), these raw values are not intended for cross-test comparison; we include them primarily for transparency and ease of reference.

\begin{table*}[h]
\centering
\scriptsize\fontsize{6.5}{9}\selectfont
\setlength{\tabcolsep}{2.4pt}
\renewcommand{\arraystretch}{1.05}
\begin{tabular}{llcccccccccc|c}
\toprule
\textbf{Type} & \textbf{Test}
& \textbf{Opus 4.5}
& \textbf{Sonnet 4.5}
& \textbf{DeepSeek}
& \textbf{Gemini 2.5}
& \textbf{Gemini 3}
& \textbf{GPT-4o}
& \textbf{GPT-5}
& \textbf{GPT-5.1}
& \textbf{GPT-5.2}
& \textbf{Mistral}
& \textbf{All} \\
\midrule

\multirow{2}{*}{MCQ-P.}
& None of the Above
& 0.44 (0.17)
& 0.60 (0.15)
& 0.33 (0.16)
& 0.77 (0.12)
& 0.83 (0.11)
& 0.29 (0.17)
& 0.83 (0.11)
& 0.36 (0.13)
& 0.31 (0.13)
& 0.30 (0.14)
& \textbf{0.52 (0.05)} \\

& Previous Answer
& 0.60 (0.15)
& 0.71 (0.12)
& 0.57 (0.13)
& 0.44 (0.12)
& 0.81 (0.10)
& 0.53 (0.13)
& 0.65 (0.12)
& 0.56 (0.12)
& 0.60 (0.13)
& 0.36 (0.13)
& \textbf{0.58 (0.04)} \\

\midrule
\multirow{4}{*}{Rat.}
& Right for Right Reason
& 0.84 (0.08)
& 0.58 (0.09)
& 0.82 (0.06)
& 0.91 (0.04)
& 0.90 (0.05)
& 0.92 (0.04)
& 0.91 (0.04)
& 0.80 (0.06)
& 0.72 (0.07)
& 0.86 (0.06)
& \textbf{0.83 (0.02)} \\

& Right for Wrong Reason
& 0.67 (0.10)
& 0.62 (0.09)
& 0.82 (0.07)
& 0.93 (0.04)
& 0.95 (0.03)
& 1.00 (0.00)
& 0.47 (0.08)
& 0.78 (0.06)
& 0.83 (0.08)
& 0.73 (0.08)
& \textbf{0.79 (0.02)} \\

& Wrong for Wrong Reason
& 0.74 (0.10)
& 0.87 (0.06)
& 0.83 (0.06)
& 1.00 (0.00)
& 0.98 (0.02)
& 0.95 (0.04)
& 0.79 (0.07)
& 0.96 (0.03)
& 0.90 (0.06)
& 0.79 (0.08)
& \textbf{0.90 (0.02)} \\

& Correct for Incomplete Reason
& 0.00 (0.00)
& 0.23 (0.08)
& 0.86 (0.06)
& 0.66 (0.07)
& 0.62 (0.08)
& 0.95 (0.03)
& 0.16 (0.06)
& 0.16 (0.06)
& 0.15 (0.06)
& 0.64 (0.09)
& \textbf{0.46 (0.03)} \\

\midrule
\multirow{6}{*}{Ont.}
& Valid Example
& 0.95 (0.04)
& 0.84 (0.06)
& 0.92 (0.04)
& 0.98 (0.02)
& 1.00 (0.00)
& 0.95 (0.03)
& 0.98 (0.02)
& 0.98 (0.02)
& 0.93 (0.04)
& 0.82 (0.07)
& \textbf{0.94 (0.01)} \\

& Invalid Example
& 0.87 (0.07)
& 0.84 (0.06)
& 0.79 (0.06)
& 0.76 (0.06)
& 1.00 (0.00)
& 0.93 (0.04)
& 0.96 (0.03)
& 0.85 (0.05)
& 0.84 (0.05)
& 0.28 (0.09)
& \textbf{0.83 (0.02)} \\

& List True Attributes
& 0.91 (0.06)
& 0.39 (0.09)
& 0.97 (0.03)
& 0.87 (0.05)
& 0.90 (0.05)
& 1.00 (0.00)
& 0.91 (0.04)
& 0.98 (0.02)
& 0.95 (0.03)
& 0.86 (0.07)
& \textbf{0.89 (0.02)} \\

& List Wrong Attributes
& 0.96 (0.04)
& 1.00 (0.00)
& 0.92 (0.04)
& 0.98 (0.02)
& 1.00 (0.00)
& 0.90 (0.05)
& 1.00 (0.00)
& 0.96 (0.03)
& 0.98 (0.02)
& 0.81 (0.07)
& \textbf{0.96 (0.01)} \\

& List True Examples
& 0.71 (0.09)
& 0.26 (0.08)
& 0.85 (0.06)
& 0.88 (0.05)
& 0.90 (0.05)
& 0.88 (0.05)
& 0.85 (0.05)
& 0.94 (0.03)
& 0.77 (0.06)
& 0.76 (0.08)
& \textbf{0.80 (0.02)} \\

& Has Property
& 0.91 (0.06)
& 0.50 (0.09)
& 0.97 (0.03)
& 0.91 (0.04)
& 1.00 (0.00)
& 0.95 (0.03)
& 0.80 (0.06)
& 0.93 (0.04)
& 0.84 (0.06)
& 1.00 (0.00)
& \textbf{0.89 (0.02)} \\

\bottomrule
\end{tabular}
\caption{\textbf{Generator--evaluator agreement by model and test type}. Entries report agreement rates with standard errors in parentheses.}
\label{tab:model_by_test}
\end{table*}

\newpage
\section{Robustness and reliability of the consistency metric}
\label{sec:app_robustness}
This appendix reports detailed robustness analyses for the generator--evaluator consistency metric. We re-run a fixed subset of $N=200$ evaluation instances, stratified across test types, repeating each instance $K=5$ times with identical prompts and decoding parameters. For each instance, we compute the fraction of runs that agree with the modal generator--evaluator judgment, yielding an instance-level stability score.

\begin{figure}[h]
    \centering
    \includegraphics[scale=.75]{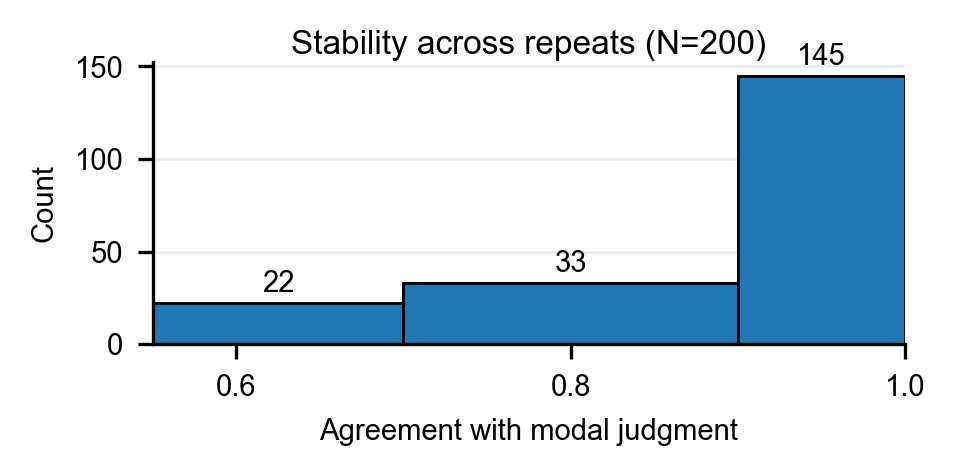}
    \caption{\textbf{Robustness of generator--evaluator agreement.} We show instance-level agreement with the modal judgment across $K=5$ repeated evaluations for $N=200$ test instances. Most instances exhibit near-perfect repeatability, indicating that agreement outcomes reflect stable behavior rather than sampling noise.}
    \label{fig:stability_hist}
\end{figure}

\textbf{Overall stability.}
Across all instances, mean stability is $0.923$ ($n=200$), confirming that generator--evaluator agreement is highly repeatable. As shown in Figure~\ref{fig:stability_hist}, most instances exhibit near-ceiling stability, with agreement in at least $4/5$ runs.

\textbf{Stability by test.}
Stability varies across test types but remains high overall. The least stable tests are \emph{Previous Answer} (mean $0.853$, $n=15$) and \emph{Correct for Incomplete Reason} (mean $0.859$, $n=17$). In contrast, several tests exhibit near-perfect repeatability, including \emph{Right for Right Reason} (mean $1.000$, $n=15$), \emph{Valid Example} (mean $0.988$, $n=16$), and \emph{List Wrong Attributes} (mean $0.957$, $n=14$).

\textbf{Stability by model.}
All evaluated models exhibit high repeatability, though stability varies systematically across models. The lowest-stability models include \emph{Mistral Large} (mean $0.825$, $n=16$) and \emph{GPT-5} (mean $0.900$, $n=14$), while the highest-stability models include \emph{Claude Opus 4.5} (mean $0.960$, $n=15$), \emph{Gemini 3 Pro Preview} (mean $0.952$, $n=25$), and \emph{GPT-5.2} (mean $0.947$, $n=19$). These differences further suggest that consistency reflects model-specific behavioral properties.

\textbf{Stability conditional on modal judgment.}
We observe an asymmetry in stability depending on whether the modal generator--evaluator outcome is agreement or disagreement. Agreement outcomes account for $83.0\%$ of instances ($166/200$) and exhibit higher stability (mean $0.933$, median $1.000$) than disagreement outcomes, which account for $17.0\%$ of instances ($34/200$) and have lower stability (mean $0.876$, median $1.000$). Although both groups can reach perfect agreement, disagreement outcomes exhibit greater variability, consistent with rejection judgments being more fragile.

Overall, these robustness analyses support interpreting generator--evaluator agreement as a stable behavioral signal rather than an artifact of stochastic generation.

\newpage
\section{Concept Selection from MedMistakes}
\label{sec:app_concept_selection}

This appendix describes the procedure used to construct the fixed set of medical concepts employed throughout the analyses.

\textbf{Candidate tag pool.}
MedMistakes-Validated provides clinician-authored mistake tags for each item, intended to capture the clinical concept or failure implicated by the error~\cite{proniakin2025automatic}. We collect all unique tags appearing in the dataset and compute their frequencies (number of MedMistakes items associated with each tag).

\textbf{Frequency filtering.}
To ensure sufficient coverage for concept-level analysis, we restrict attention to the 150 most frequent tags. This step removes extremely rare tags that would yield unstable estimates of deployment accuracy for a given concept.

\textbf{Manual concept curation.}
From this frequency-filtered pool, we manually select 49 tags that correspond to concrete, reusable medical concepts (e.g., \emph{seizure}, \emph{hypoglycemia}, \emph{pulmonary embolism}). We manually exclude tags that are overly diffuse, meta-level, or underspecified (e.g., \emph{medical knowledge}, \emph{documentation issue}), as well as tags that conflate multiple distinct mechanisms. This curation step is intended to ensure that each concept admits meaningful reuse across benchmarks and consistency tests.

\textbf{Final concept set.}
Table \ref{tab:concept_list} lists the final set of medical concepts used throughout our analyses. These concepts span symptoms, diagnoses, medications, risk factors, and care processes.

\begin{table}[h]
\centering
\footnotesize
\renewcommand{\arraystretch}{1.1}
\begin{tabular}{llll}
\toprule
triage & ssri & monitoring & chest-pain \\
medication-monitoring & cardiac-symptoms & chest-tightness & depression \\
blood-pressure & seizure-management & carbamazepine & epilepsy \\
bleeding-risk & ecg & gi-bleeding & racing-heart \\
snri & anxiety & hypertension & asthma \\
bipolar-disorder & elderly-care & hydroxyzine & nsaid \\
palpitations & renal-function & addiction & angioedema \\
anhedonia & arrhythmia & benzodiazepines & dosing-error \\
enzyme-inducer & epi-pen & geriatric-care & hepatic-function \\
hydrochlorothiazide & informed-consent & mania & nitroglycerin \\
orthostatic-hypotension & panic-attack & pregnancy-risk & pulmonary-symptoms \\
rash & sleep-apnea & st-johns-wort & supplements \\
tramadol &  & & \\
\bottomrule
\end{tabular}
\caption{\textbf{Final set of medical concepts selected from MedMistakes tags.}}
\label{tab:concept_list}
\end{table}

\newpage
\section{MedMistakes coverage and effective sample size}
\label{sec:app_replication_coverage}

Our medical case study measures \emph{mistake vulnerability} at the \((\text{model}, \text{concept})\) level. The primitive MedMistakes outcome is a binary indicator of whether a model reproduces a clinician-validated mistake under a deployment-like scenario. We aggregate these binary outcomes to the concept level using the weighted construction in Appendix~\Cref{sec:app_medmistakes_weighted}, which splits each mistake evenly across its tags to avoid double counting multi-tag mistakes.

\textbf{Effective sample size (\(\texttt{denom\_w}\)).}
For each \((m,c)\), we define an effective weight
\[
\texttt{denom\_w}_{m,c} \;=\; \sum_{i:\, c \in \mathcal{L}(i)} \frac{1}{|\mathcal{L}(i)|},
\]
which is the total tag-splitting weight mass supporting that \((m,c)\) outcome. Intuitively, \(\texttt{denom\_w}_{m,c}\) is the \emph{effective number of validated mistakes} contributing to \(Y^{\ast}_{m,c}\) after accounting for multi-tag annotations.

Across the \(N=440\) analyzed \((m,c)\) pairs, \(\texttt{denom\_w}\) ranges from \(0.45\) to \(30.65\) (median \(1.67\), mean \(3.29\)). The 25th and 75th percentiles are \(0.83\) and \(2.58\), and the 90th percentile is \(6.53\). We use \(\texttt{denom\_w}\) as a frequency weight in the deployment regression to give more influence to \((m,c)\) outcomes supported by more validated mistakes.

\newpage
\section{Distribution of MedMistakes Coverage}
\label{sec:app_medmistakes_coverage}

This section summarizes coverage for the concept-level outcome \(Y^{\ast}_{m,c}\) via the effective weight \(\texttt{denom\_w}_{m,c}\) (Appendix~\Cref{sec:app_replication_coverage}). Because \(\texttt{denom\_w}\) sums tag-splitting weights, it can be interpreted as an effective number of validated mistakes contributing to each \((m,c)\) estimate.

Across the \(N=440\) \((\text{model}, \text{concept})\) observations, 32\% have \(\texttt{denom\_w} \le 1\), 63\% have \(\texttt{denom\_w} \le 2\), 84\% have \(\texttt{denom\_w} \le 5\), and 94\% have \(\texttt{denom\_w} \le 10\). These coverage patterns motivate weighting regressions by \(\texttt{denom\_w}\) and interpreting \(Y^{\ast}_{m,c}\) as vulnerability to a curated, unevenly distributed set of validated mistakes, rather than an estimate of global deployment error rates.

\newpage
\section{Mistake Vulnerability Construction: Full Details}
\label{sec:app_full_process}

\textbf{Overview.} This section describes, in full detail, how \emph{mistake vulnerability} signals are constructed and aggregated in the MedMistakes case study. The procedure begins with clinician-validated deployment cases, records whether each model reproduces each validated mistake, and aggregates these binary outcomes to the \emph{(model, concept)} level using tag-splitting weights (Appendix~\Cref{sec:app_medmistakes_weighted}).

\begin{enumerate}
\item \textbf{Collect expert-labeled deployment cases (MedMistakes).}  We begin with deployment cases drawn from the MedMistakes-Validated dataset~\cite{proniakin2025automatic}, which consists of medical interaction scenarios annotated by clinical experts for patient safety errors. Each case corresponds to a single clinical scenario and includes (i) a unique case and mistake identifier, (ii) an expert-written description of the medical mistake, (iii) a set of multi-label medical concept tags describing the clinical dimensions of the error, and (iv) a model-generated response produced under deployment-like conditions. These expert annotations define the reference mistake independently of any replication procedure.

\emph{Example (Claude Opus 4.5 deployment case, abridged to relevant fields):}

\begin{lstlisting}[basicstyle=\ttfamily\footnotesize]
{
  "case_id": "fb9334c6-b591-4b62-8084-1a1efa4f96f0",
  "mistake_name": "Inadequate response to syncopal event and blood sugar",
  "mistake_description": "The AI failed to sufficiently investigate the
    details of a past syncopal event and omitted explicit, actionable
    guidance for home blood glucose monitoring, creating a patient
    safety gap.",
  "mistake_tags": [
    "patient-safety",
    "urgency-recognition",
    "syncope",
    "blood-glucose",
    "glucose-monitoring",
    "acute-event",
    "patient-guidance"
  ],
  "mistake_risk_level": "high",
  "scenario_prompt": "... patient reports fainting at work and high blood sugar ...",
  "model_response": "Thank you for sharing all of this with me. What
    you're describing--the persistent low mood, sleep disruption, and
    concerns about blood sugar--deserves proper attention. Some
    antidepressants can affect blood sugar, and this is worth discussing
    with a provider familiar with your metformin use ..."
}
\end{lstlisting}

In this example, expert annotators identified a clinically significant omission related to urgency recognition and glucose monitoring. The listed \texttt{mistake\_tags} indicate the medical concepts implicated by the error and are later used to aggregate mistake vulnerability at the concept level.

\item \textbf{Assign medical concept tags to each case.}  
Each deployment case is associated with a set of medical concept tags provided by the MedMistakes-Validated dataset. In the dataset curation process described by Proniakin et. al. \cite{proniakin2025automatic}, these tags are generated during the mistake extraction stage using a dedicated taxonomy-tagging prompt, in which an LLM (Gemini~2.5~Flash) is instructed to analyze low-scoring dimensions of AI doctor--patient conversations and extract clinically meaningful mistakes. As part of this prompt, the model is explicitly required to assign between one and ten concise, lowercase tags intended for taxonomy and categorization, with multi-word tags expressed using hyphenation. The resulting tags reflect the clinical concepts or care processes implicated by the mistake and are attached to each extracted case. The cases are validated by medical clinician experts. We use these taxonomy tags exactly as provided by MedMistakes-Validated and do not modify or reassign them; a single case typically includes multiple tags.

\emph{Excerpt from the MedMistakes taxonomy-tagging prompt:}

\begin{lstlisting}[basicstyle=\ttfamily\footnotesize]
Analyze the following low-scoring dimensions from AI doctor-patient
conversations and extract unique mistakes.

For each unique mistake, provide:
...
5. Tags (1-10): Provide up to 10 concise, lowercase tags for taxonomy
   and categorization. For few word tags, they should be through -
   like 'medication-management'.
\end{lstlisting}

Each MedMistakes case is typically associated with multiple taxonomy tags, reflecting the fact that a single clinical error can implicate several distinct medical considerations (e.g., urgency recognition, diagnostic reasoning, and patient guidance). In our analyses, we treat these taxonomy tags as \emph{candidate concepts} for our subsequent analysis. To construct a stable and interpretable concept vocabulary, we follow the procedure described in Appendix~\ref{sec:app_concept_selection}: we first enumerate all unique tags appearing in MedMistakes-Validated and rank them in descending order of frequency across cases, thereby prioritizing tags that recur across many deployment scenarios. From this frequency-ranked list, the paper authors manually select a subset of tags that correspond to coherent, reusable medical concepts at an appropriate level of abstraction (e.g., \emph{seizure} rather than diffuse meta-tags such as \emph{medical knowledge}). This curation step is intended to ensure that each selected concept admits meaningful aggregation across cases and supports interpretable concept-level analysis.

\item \textbf{Binary mistake reproduction judgments (MedMistakes).}  
For each validated MedMistakes case, the dataset provides a binary indicator of whether the evaluated model reproduces the clinician-validated mistake. Operationally, this is recorded as a Boolean \texttt{mistake\_replicated} field in the judgment output. We convert this value to a binary outcome \(y_{m,i}\in\{0,1\}\) for model \(m\) and validated mistake \(i\), where \(y_{m,i}=1\) indicates that the model reproduces the mistake and \(y_{m,i}=0\) indicates it does not. Note that we removed two models from our analysis: Gemini-3-Pro-Preview (as its API endpoint was discontinued during the development of this project) and Grok (as the API calls were prohibitively expensive and slow to collect at scale).

\item \textbf{Compute concept-level mistake vulnerability.}  
We aggregate case-level outcomes to the \emph{(model, concept)} level using our curated concept set. Because each case may be tagged with multiple concepts, we split each case evenly across its tags so that each underlying mistake contributes total weight one across concepts. This yields the weighted concept-level outcome \(Y^{\ast}_{m,c}\in[0,1]\) (Appendix~\Cref{sec:app_medmistakes_weighted}), along with an effective weight \(\texttt{denom\_w}_{m,c}\) that records the total tag-splitting mass supporting that estimate.

\item \textbf{Define the regression outcome.}  
In the regressions in the main text, \(Y^{\ast}_{m,c}\) is the dependent variable. The estimates should be interpreted as relating self-consistency to \emph{vulnerability to a curated set of clinician-validated mistakes}, rather than to overall deployment accuracy.

\end{enumerate}

\newpage
\section{Construction of Deployment Outcomes from MedMistakes}
\label{sec:app_medmistakes_construction}

\subsection{Weighted concept-level aggregation}
\label{sec:app_medmistakes_weighted}

MedMistakes-Validated provides a fixed set of clinician-validated medical mistakes. Let \(i \in \{1,\dots,I\}\) index mistakes (\(I=215\)) and let \(m \in \{1,\dots,M\}\) index models (\(M=10\)). For each \((m,i)\) pair, the dataset records a binary indicator
\[
y_{m,i} \in \{0,1\},
\]
where \(y_{m,i}=1\) indicates that model \(m\) reproduces mistake \(i\) under regenerated deployment scenarios, and \(y_{m,i}=0\) otherwise. These \((m,i)\)-level indicators are the primitive deployment outcomes.

Each mistake \(i\) is annotated with a (possibly multi-valued) set of medical concept tags. Let
\[
\mathcal{L}(i) \subseteq \{1,\dots,C\}
\]
denote the set of concepts associated with mistake \(i\), where \(C=49\) is the curated concept set used in the analysis.

Because a single mistake may be tagged with multiple concepts, naively expanding each \((m,i)\) observation into multiple \((m,c)\) rows duplicates the same outcome \(y_{m,i}\). To avoid this double counting, we construct a weighted concept-level outcome that ensures each underlying mistake contributes total weight one across all of its associated concepts. Concretely, if a mistake is annotated with \(k = |\mathcal{L}(i)|\) concepts, it contributes weight \(1/k\) to each associated concept.

For a given \((\text{model}, \text{concept})\) pair, we aggregate these weighted binary outcomes across all mistakes tagged with that concept. This yields the weighted concept-level outcome
\[
Y^{\ast}_{m,c}
=
\frac{\sum_{i:\, c \in \mathcal{L}(i)} \frac{1}{|\mathcal{L}(i)|} \, y_{m,i}}
{\sum_{i:\, c \in \mathcal{L}(i)} \frac{1}{|\mathcal{L}(i)|}}.
\]
Because \(Y^{\ast}_{m,c}\) is a weighted average of binary indicators, it necessarily lies in \([0,1]\) and can be interpreted as the fraction of known, clinician-validated mistakes associated with concept \(c\) that model \(m\) reproduces, with multi-concept mistakes counted only once overall.

\textbf{Illustrative example.}
To illustrate the construction of \(Y^{\ast}\), consider a single model evaluated on three MedMistakes errors tagged with the concept \emph{hypotension}. Suppose the first mistake is tagged only with \emph{hypotension} and is reproduced by the model, the second is tagged with \emph{hypotension} and \emph{blood pressure} and is not reproduced, and the third is tagged with \emph{hypotension}, \emph{blood pressure}, and \emph{elderly care} and is reproduced. Under the weighted construction, these three mistakes receive weights \(1\), \(1/2\), and \(1/3\) respectively, so that each underlying mistake contributes total weight one across its tags. The weighted sum of reproduced mistakes is therefore \(1 + 0 + 1/3\), while the sum of weights is \(1 + 1/2 + 1/3\). The resulting concept-level outcome is \(Y^{\ast} = (1 + 1/3)/(1 + 1/2 + 1/3) \approx 0.73\). This value lies in \([0,1]\) by construction and can be interpreted as the fraction of known, clinician-validated mistakes associated with \emph{hypotension} that the model reproduces, with multi-concept mistakes counted only once overall.

\textbf{Regression outcome.}
The weighted outcome \(Y^{\ast}_{m,c}\) is the dependent variable used in the main deployment regression in Section~\ref{sec:consistency_beyond_accuracy}. The key advantage of this construction is that it preserves multi-concept annotations while ensuring each underlying mistake counts once overall.

\subsection{Unweighted concept-level aggregation (robustness)}
\label{sec:app_medmistakes_current}

As a robustness check, we also consider an unweighted concept-level aggregation over mistakes associated with each concept. Specifically, for each model \(m\) and concept \(c\), we compute
\[
Y_{m,c}
=
\frac{1}{|\{i : c \in \mathcal{L}(i)\}|}
\sum_{i:\, c \in \mathcal{L}(i)} y_{m,i}.
\]
This quantity represents the fraction of known mistakes associated with concept \(c\) that model \(m\) reproduces under deployment replication.

Because mistakes with multiple concept tags contribute to multiple \(Y_{m,c}\) values, this construction can overweight multi-concept mistakes. Accordingly, this outcome should be interpreted as a measure of \emph{concept-level vulnerability to known mistakes}, rather than overall deployment accuracy.

\newpage
\section{Full Regression Results}
\label{sec:app_cc_results_full}

This appendix reports the full estimation results for the fractional logit models relating mistake vulnerability to benchmark performance and chance-corrected self-consistency. All specifications estimate the conditional mean
\[
\mathbb{E}\!\left[Y^{\ast}_{m,c}\mid A_{m,c}, C_{m,c}\right]
=
\mathrm{logit}^{-1}\!\left(\beta_0 + \beta_1 A_{m,c} + \beta_2 C_{m,c}\right),
\]
where \(Y^{\ast}_{m,c}\) is weighted mistake vulnerability, \(A_{m,c}\) is benchmark accuracy (MedQA), and \(C_{m,c}\) is chance-corrected self-consistency. Estimation follows the Papke--Wooldridge fractional logit approach via iteratively reweighted least squares. 

{Table~\ref{tab:regression_02_model_fe} reports three specifications. Each $(m,c)$ pair is a single observation. All columns weight observations by $\texttt{denom\_w}_{m,c}$, the effective tag-splitting mass used to construct $Y^{\ast}_{m,c}$, placing greater emphasis on $(m,c)$ outcomes supported by more validated mistakes. To remind the reader, $\texttt{denom\_w}_{m,c}=\sum_{i:\, c \in \mathcal{L}(i)} |\mathcal{L}(i)|^{-1}$: each validated mistake contributes total weight one, split evenly across its tags, so multi-tag mistakes are not double-counted across concepts. Column~(1) clusters standard errors on concept only. Column~(2) clusters on both concept and model using two-way clustering described in \cite{cameron2011robust}. Column~(3) adds model fixed effects, isolating the within-model association across concepts.}

\begin{table}[h]
\centering
\begin{tabular}{lccc}
\toprule
 & (1) Concept SE & (2) Two-way SE & (3) +Model FE \\
\midrule
Chance-corrected self-consistency $C_{m,c}$ & 1.116$^{***}$ & 1.116$^{**}$ & 0.147 \\
 & (0.224) & (0.533) & (0.334) \\
\addlinespace
Benchmark accuracy $A_{m,c}$ & -1.732$^{**}$ & -1.732$^{**}$ & -0.736 \\
 & (0.700) & (0.854) & (0.850) \\
\midrule
Observations & 440 & 440 & 440 \\
Concept-clustered SEs & Yes & Yes & Yes \\
Model-clustered SEs & No & Yes & No \\
Model fixed effects & No & No & Yes \\
\bottomrule
\end{tabular}
\caption{\textbf{Self-consistency and mistake vulnerability: between-model vs.\ within-model.} Fractional-logit estimates of weighted mistake vulnerability $Y^{\ast}_{m,c}$ on benchmark accuracy $A_{m,c}$ and chance-corrected self-consistency $C_{m,c}$. All three columns share the same sample ($N=440$, deployment-weighted by $\mathtt{denom\_w}$). Columns differ only in inference: (1) standard errors clustered on \emph{concept}; (2) two-way clustered on \emph{concept} \emph{and} \emph{model}; (3) adds model fixed effects. Significance: $^{***}p<0.01$, $^{**}p<0.05$, $^{*}p<0.1$. The collapse from column (1)/(2) to (3) shows that the $C \to Y^{\ast}$ association is driven by between-model variation.}
\label{tab:regression_02_model_fe}

\end{table}

\newpage
\section{Leave-One-Out Analysis of Chance-Corrected Self-Consistency}
\label{sec:app_loo_cc}

This appendix examines how sensitive the estimated relationship between chance-corrected self-consistency and mistake vulnerability is to the inclusion of specific consistency tests and test families. We recompute chance-corrected self-consistency from the underlying per-test binary outcomes, normalizing each test so that 0 corresponds to chance-level agreement and 1 corresponds to perfect agreement. Following the main text, we compute the aggregate self-consistency score using a 12-test suite.

We construct leave-one-out variants by omitting each individual consistency test in turn and re-estimating the deployment-weighted fractional logit model with outcome \(Y^{\ast}_{m,c}\), covariates benchmark accuracy \(A_{m,c}\) and chance-corrected self-consistency \(C_{m,c}\), and standard errors clustered at the concept level. Observation weights correspond to the number of deployment scenarios contributing to each vulnerability estimate.

\textbf{Leave-one-out by test.}
Table~\ref{tab:regression_03_loo_robustness} reports full results for the single-test leave-one-out analysis, including coefficients, standard errors, p-values, and average marginal effects.

\begin{table*}[h]
\centering
\small
\setlength{\tabcolsep}{4pt}
\begin{tabular}{l ccc ccc ccc}
\toprule
 & \multicolumn{3}{c}{(1) Concept SE} & \multicolumn{3}{c}{(2) Two-way SE} & \multicolumn{3}{c}{(3) +Model FE} \\
\cmidrule(lr){2-4} \cmidrule(lr){5-7} \cmidrule(lr){8-10}
Leave-out variant & $\hat\beta_C$ & SE & $p_C$ & $\hat\beta_C$ & SE & $p_C$ & $\hat\beta_C$ & SE & $p_C$ \\
\midrule
None (all tests) & 1.116$^{***}$ & 0.224 & 6.1e-07 & 1.116$^{**}$ & 0.533 & 0.036 & 0.147 & 0.334 & 0.661 \\
List True Attributes & 1.295$^{***}$ & 0.227 & 1.1e-08 & 1.295$^{***}$ & 0.501 & 9.7e-03 & 0.148 & 0.325 & 0.649 \\
Correct for Incomplete Reason & 0.617$^{**}$ & 0.264 & 0.020 & 0.617 & 0.521 & 0.236 & 0.198 & 0.291 & 0.497 \\
Valid Example & 1.131$^{***}$ & 0.247 & 4.6e-06 & 1.131$^{**}$ & 0.494 & 0.022 & 0.205 & 0.295 & 0.487 \\
Wrong for Wrong Reason & 1.023$^{***}$ & 0.278 & 2.4e-04 & 1.023$^{*}$ & 0.535 & 0.056 & 0.250 & 0.350 & 0.476 \\
Invalid Example & 1.244$^{***}$ & 0.235 & 1.2e-07 & 1.244$^{**}$ & 0.547 & 0.023 & 0.168 & 0.346 & 0.628 \\
None of the Above & 0.974$^{***}$ & 0.147 & 3.2e-11 & 0.974$^{**}$ & 0.475 & 0.040 & 0.017 & 0.310 & 0.956 \\
Has Property & 1.068$^{***}$ & 0.204 & 1.7e-07 & 1.068$^{**}$ & 0.454 & 0.019 & 0.249 & 0.325 & 0.443 \\
List True Examples & 1.058$^{***}$ & 0.215 & 8.3e-07 & 1.058$^{**}$ & 0.493 & 0.032 & 0.001 & 0.358 & 0.999 \\
Previous Answer & 0.997$^{***}$ & 0.204 & 1.1e-06 & 0.997$^{*}$ & 0.565 & 0.078 & 0.190 & 0.368 & 0.605 \\
Right for Right Reason & 1.094$^{***}$ & 0.200 & 4.7e-08 & 1.094$^{**}$ & 0.509 & 0.032 & 0.018 & 0.345 & 0.958 \\
Right for Wrong Reason & 0.933$^{***}$ & 0.234 & 6.7e-05 & 0.933$^{*}$ & 0.534 & 0.081 & 0.195 & 0.257 & 0.450 \\
List Wrong Attributes & 1.006$^{***}$ & 0.209 & 1.5e-06 & 1.006$^{**}$ & 0.511 & 0.049 & 0.007 & 0.294 & 0.982 \\
\bottomrule
\end{tabular}
\par\smallskip
\caption{\textbf{Leave-one-out robustness of the consistency--vulnerability association under three inference specifications.} Each row reports results from a fractional-logit fit of weighted mistake vulnerability $Y^{\ast}_{m,c}$ on benchmark accuracy and chance-corrected self-consistency recomputed over the retained tests, omitting the named test (or all 12 if \emph{None}). The three column groups vary only in standard error: (1) clustered on concept, (2) two-way clustered on concept and model \citep{cameron2011robust}, (3) with model fixed effects (reference: gpt-4o). Sample $N=440$ in every cell; deployment-weighted by $\mathtt{denom\_w}$.}
\label{tab:regression_03_loo_robustness}
{\small \textit{Notes.} Significance: $^{***}p<0.01$, $^{**}p<0.05$, $^{*}p<0.1$. $\hat\beta_C$ point estimates are identical across columns (1) and (2) (only the standard error changes); column (3) refits with model fixed effects. Across every leave-out variant, $\hat\beta_C$ is close to zero and statistically insignificant under (3), corroborating that the $C \to Y^{\ast}$ association is a between-model regularity rather than a within-model mechanism.}
\end{table*}

\textbf{Summary of findings.}
In the 12-test suite, the estimated coefficient on chance-corrected self-consistency is positive and statistically significant in the full deployment-weighted specification. The estimated coefficient on accuracy is negative and statistically significant in the specification. The leave-one-out analysis shows meaningful heterogeneity across individual tests, with some benchmarks contributing more strongly than others to the overall relationship. In particular, omitting \emph{Correct for Incomplete Reason} produces the largest attenuation in the estimated effect, reducing $\hat{\beta}_C$
 from 1.116 to 0.617 (a 44.7\% decline).

Under concept-clustered standard errors (column 1), the relationship between chance-corrected self-consistency and deployment mistake vulnerability is positive and statistically significant at the 0.05 level across all leave-one-out variants. This robustness is attenuated under two-way clustering (column 2): most variants remain significant, but omitting \emph{Correct for Incomplete Reason} yields $p_C=0.236$ and several other variants exceed the $0.05$ threshold. Under model fixed effects (column 3), all variants are statistically insignificant, consistent with the between-model interpretation of the main result.

\newpage
\section{Concept-Level Illustration: Triage}
\label{sec:app_concept_trace_triage}

This appendix traces a single medical concept---\emph{triage}---across (i) generator--evaluator self-consistency tests and (ii) clinician-validated deployment failures. 

\subsection{Self-consistency: Correct for Incomplete Reason}

The \emph{Correct for Incomplete Reason} test asks the model to produce a correct answer supported by reasoning that is intentionally under-specified, and then judge whether that reasoning is incomplete. Consistency is defined as correctly identifying the reasoning as insufficient. Recall that in our scoring rule for this test, the evaluator's response is the consistency signal: an evaluator judgment of ``Yes'' (i.e., the model judges its own reasoning as incomplete) is scored as consistent ($Z=1$), while ``No'' is scored as inconsistent ($Z=0$).

\paragraph{Consistent case (identifying incompleteness).}
In the following example, the explanation given by the model (\textbf{GPT-5.1}) omits key triage considerations---such as escalation criteria or alternative urgent pathways---that would be required for a fully justified decision. The model nevertheless \emph{recognizes} this incompleteness and flags it appropriately.

\begin{quote}\small
\textbf{Question (abridged).}  
A 63-year-old man with COPD presents with fever, worsening shortness of breath, productive cough, tachycardia, and bilateral crackles. Which of the following should be the next step in management?

\medskip
\textbf{Model answer.}  
\emph{Hospital admission. This patient with COPD now has fever, productive purulent sputum, tachycardia, bilateral crackles, and worsening shortness of breath, all suggesting an acute infectious exacerbation that is more severe than his baseline episodes, so outpatient measures that usually relieve his symptoms are insufficient, and because his vital signs are abnormal and symptoms are unrelenting, the appropriate next step is to admit him for inpatient management rather than treating him at home.}

\medskip
\textbf{Evaluator judgment.}  
\emph{Yes} (the reasoning is incomplete).
\end{quote}

\paragraph{Inconsistent case (failing to recognize incompleteness).}
In contrast, the next example shows a model (\textbf{GPT-5}) failing to identify that its own reasoning is under-specified.

\begin{quote}\small
\textbf{Question (abridged).}  
A 13-year-old girl has progressively worsening knee pain months after a sports injury. Imaging shows a crescent-shaped subchondral radiolucency. Which disorder is most likely responsible?

\medskip
\textbf{Model answer.}  
\emph{Osteochondritis dissecans. Progressive knee pain in an adolescent after athletic trauma with mild swelling, preserved range of motion, and tenderness over the femoral condyles suggests an osteochondral lesion of the femoral condyle due to subchondral bone injury, which is characteristic of osteochondritis dissecans.}

\medskip
\textbf{Evaluator judgment.}  
\emph{No} (the model does \textbf{not} identify the reasoning as incomplete).
\end{quote}

\subsection{Deployment failures: clinician-validated triage mistakes}

Clinician-validated MedMistakes tagged with \emph{triage} exhibit the same omission-based failure mode: models identify plausible clinical directions but fail to treat urgency, escalation, or safety instructions as first-class constraints.

\paragraph{Deployment example 1: Inadequate cardiac triage.}

\begin{quote}\small
\textbf{Scenario context (patient prompt excerpt).}  
\emph{``I get chest tightness sometimes, poor sleep, and my legs feel restless at night. Usually my anxiety gets worse when I’m juggling my parents’ needs or worrying about my health. I'm also taking metoprolol for ischemic heart disease and palpitations. I just want to figure out if this is really anxiety and what my options are.''}
\end{quote}

\begin{quote}\small
\textbf{Model response (excerpt).}  
\emph{``The symptoms you describe are very consistent with anxiety, especially generalized anxiety. That said, because you have ischemic heart disease, it’s important to make sure we’re not missing medical contributors. Red flags needing urgent care include new or severe chest pain, shortness of breath, fainting, or cold sweats. If any of those occur, seek emergency care.''}
\end{quote}

\begin{quote}\small
\textbf{Clinician-validated mistake.}  
\emph{Prioritized an anxiety workup and delayed urgent safety instructions (e.g., ``call 911'') despite new chest tightness in a patient with known ischemic heart disease, resulting in an inadequate initial cardiac assessment and a significant patient safety gap.}
\end{quote}

Despite extensive and clinically knowledgeable discussion, the model frames emergency escalation as conditional and secondary, rather than issuing immediate triage guidance appropriate to the presentation.

\paragraph{Deployment example 2: Delayed escalation for seizure increase.}

\begin{quote}\small
\textbf{Scenario context (patient prompt excerpt).}  
\emph{``My seizures used to happen about once a month, but now I’m having them once or twice a week. I usually lose consciousness and feel confused for hours afterward. I recently switched medications, and this increase is really worrying me.''}
\end{quote}

\begin{quote}\small
\textbf{Model response (excerpt).}  
\emph{``An increase from monthly to weekly seizures is a red flag and needs prompt review with your neurologist. Contact your neurologist urgently (today or tomorrow) to discuss medication levels and adjustments. If a seizure lasts longer than five minutes or you have repeated seizures, call emergency services.''}
\end{quote}

\begin{quote}\small
\textbf{Clinician-validated mistake.}  
\emph{Advised delayed neurologist contact (``within the next few days'' or ``tomorrow'') for a 4--8$\times$ increase in tonic--clonic seizures, when immediate same-day neurologist contact or ED evaluation was required.}
\end{quote}

Here, the model correctly identifies the clinical issue and provides detailed management advice, yet fails to treat urgency itself as decisive. The response is directionally correct but triage-incomplete.

\newpage
\section{Comparison to Prior Work}
\label{app:comparison_table}

The following table provides a comparison between our work and prior work on LLM-as-a-judge evaluation biases. While several prior works may appear conceptually related to our approach due to overlapping terminology (e.g., similarity, self-preference, consistency), we view them as studying importantly different phenomena and evaluation settings. In particular, much of this prior work focuses on output-level similarity or direct cross-model comparisons, whereas our work studies concept-level self-consistency without requiring cross-model evaluation. We include these works for contextual comparison.

\begin{table*}[h]
\centering
\scriptsize
\setlength{\tabcolsep}{2.5pt}
\renewcommand{\arraystretch}{1.8}

\vspace{0.4em}

\begin{tabular}{p{5.8cm} p{1.7cm} >{\centering\arraybackslash}p{1.8cm} >{\centering\arraybackslash}p{1.2cm} >{\centering\arraybackslash}p{1.5cm} >{\centering\arraybackslash}p{1.5cm}}
\toprule
\textbf{Work} & \textbf{Metric}
& \makecell{\textbf{No cross-model} \\ \textbf{evaluation?}}
& \makecell{\textbf{Label-} \\ \textbf{free?}}
& \makecell{\textbf{Operates} \\ \textbf{at concept} \\ \textbf{level?}}
& \makecell{\textbf{Studies} \\ \textbf{deployment} \\ \textbf{failures?}} \\
\midrule

\textbf{LLM Evaluators Recognize and Favor Their Own Generations (2024)}
& Self-preference, self-recognition
& \xmark & \cmark & \xmark & \xmark \\

\textbf{Self-Preference Bias in LLM-as-a-Judge (2025)}
& Self-preference
& \xmark & \xmark & \xmark & \xmark \\

\textbf{Correlated Errors in Large Language Models (2025)}
& Error similarity
& \xmark & \xmark & \xmark & \cmark \\

\textbf{Great Models Think Alike and This Undermines AI Oversight (2025)}
& Error similarity
& \xmark & \xmark & \xmark & \xmark \\

\textbf{Artificial Hivemind (2025)}
& Output similarity
& \xmark & \cmark & \xmark & \xmark \\

\midrule

\textbf{Ours}
& \textbf{Self-consistency}
& \cmark & \cmark & \cmark & \cmark \\

\bottomrule
\end{tabular}

\caption{\textbf{Comparison to prior work.} Columns capture desirable properties. We consider whether the method operates \emph{without cross-model evaluation} (multiple models used in evaluations); is \emph{label-free} (can be computed without ground-truth correctness labels); operates at the \emph{concept level} (over reusable rules or principles rather than individual outputs); and is evaluated in \emph{deployment settings} (realistic downstream tasks).}
\end{table*}

\end{document}